# "Phase Transitions" in small systems: why standard threshold definitions fail for nanolasers

G.L. Lippi

*Université Côte d'Azur, CNRS, Institut de Physique de Nice*

T. Wang

*School of Electronics and Information, Hangzhou Dianzi University, Hangzhou 310018, China*[*]

G.P. Puccioni

*Istituto dei Sistemi Complessi, CNR, Via Madonna del Piano 10, I-50019 Sesto Fiorentino, Italy*

(Dated: February 12, 2022)

Since the development of micro- and nanolasers, the question of laser threshold has been subject to debate. Different definitions have been used to try and establish its occurrence, often encountering major obstacles. We examine a set of common physical definitions which we apply to measurements taken in a micro-VCSEL. Their predictions not only clearly disagree, pointing to different pump values at which the laser should cross threshold, but they also correspond to autocorrelation values which demonstrate very low field coherence. A topological analysis of the rate equations, with average spontaneous emission added to the lasing mode, clearly identifies the contradictions and explains the origin of the discrepancies. Additional considerations help understanding the failure of the approach and highlight the path towards a unique and general definition of threshold in all lasers, irrespective of their sizes. A critical scrutiny of the assumptions made in the rate equations with spontaneous emission illustrates their strength and weaknesses and better defines the bounds within which their predictions hold. We remark in the conclusions how the main results of this paper could hold for other small systems.

## I. INTRODUCTION

After more than sixty years of existence and a phenomenal development, lasers have become ubiquitous in everybody's life. Hence, the idea of a *laser threshold* – the pump rate value from which coherent emission takes hold – falls only short from becoming household terminology and is certainly deeply set into the vocabulary of anyone using lasers in a professional capacity. The way the concept is visualized may depend on the community that is using it and the multiple definitions are not necessarily strictly compatible, from a conceptual point of view, but satisfy practical requirements.

Forgetting functional aspects, one would expect the laser threshold to be a well-established and uniquely defined concept, after such a long time in existence. This is, unfortunately, not the case and has caused the longstanding debate about threshold in small devices. We are going to examine some of its common definitions and show that they are equivalent only for "*large*" lasers, i.e., what amounts to the so-called *thermodynamic limit*.

The origin of the numerous threshold definitions can be traced to the related physical interpretations, founded either on statistical concepts, energy considerations or balance arguments in radiation theory. In addition, experimental constraints – stemming from the access to measurable quantities – have contributed to introducing practical analogues (e.g., linewidth reduction) whose validity may become at times questionable. As shown in this paper, all the considered definitions predict the same thresh-old value macroscopic lasers, thereby acquiring equal status. This equivalence has introduced a good degree of confusion in the definition of threshold in smaller lasers, where the predictions differ.

The way of solving the problem rests, of course, on the identification of a proper threshold definition, then on the search for experimentally accessible signatures through which the feature can be identified. This paper aims at experimentally demonstrating, then explaining on the basis of a simple model, the differences among the *accessory definitions*; it will further point out the primary one and suggest experimental techniques for its measurement.

Laser volume reduction began earnestly at the end of the 1970s with the proposal of the Vertical Cavity Surface Emitting Laser (VCSEL) [1, 2] and was propelled by the promise for applications (cf. [3–5] for a contemporary perspective). Smaller devices offered various advantages, such as strong miniaturization, lower power consumption and small thermal load, while producing a coherent electromagnetic field which suits a very large number of applications.

The progressive miniaturization of laser cavities carried the realization that the threshold definition, based on the abrupt change in slope in the photon flux (traditional of macroscopic lasers)[1], could no longer hold and that the identification of the pump value at which threshold is crossed became ill-defined. The extrapolation to the limiting case where only one electromagnetic cavity mode for the spontaneous emission is present inside the resonator ($\beta = 1$, with $\beta$ representing the fraction of spontaneous emission coupled into the lasing mode) led to the introduction of the concept of "thresholdless laser" [6]: a device where emission *is expected* to be coherent from its very first photon. More paradoxical yet, was the denomination "zero-threshold laser" [7] which extrapolated the threshold to zero pump rate. While the latter concept was soon dispelled [8], the former one is still used to denote those devices where no appreciable change in slope appears in the input-output laser response curve[2] (cf., e.g., [10–12]).

---

[*] wangtao@hdu.edu.cn

[1] Due to the strong contribution of spontaneous emission below threshold in micro- and nanolasers, an alternative, customary, representation consists in plotting the laser response in a double logarithmic curve. In this case, theoretical models show two parallel branches (lower matching incoherent, upper representing coherent emission) where the average photon number increases with pump, and are connected, at threshold, by a steeper line [6].

Clearly, if the definition of threshold relies, as it always did for macroscopic lasers, on the presence of a sharp jump between emission branches (or in the output power slope, in linear representation), then the threshold is no longer identifiable. However, this vision is very partial and amounts to reducing the emerging complex physics, which develops around the laser threshold, to a simple, abrupt change in coherence.

Decades of work have accumulated sufficient experimental knowledge and numerical evidence to show that the transition between the emission of incoherent photons and that of a coherent macroscopic field is much more complex. While the physics which underlies this transition is not expected to change as a function of cavity volume (or $\beta$), it is only in sufficiently small devices ($\beta > 10^{-4}$) that it is possible to highlight its existence and to study its features.

This paper examines different pieces of experimental evidence, coming from one same microlaser system, and applies various threshold definitions to the measurements. The contradictory results are then critically examined by comparing them to the information that can be gathered from the simplest laser model: the Rate Equations adapted to include the (average) contribution of the Spontaneous Emission (RESE). This extends the analysis to all laser sizes. A critical scrutiny of the properties of this model, compared to those of the standard Rate Equations (REs), also shows how the different threshold definitions – which predict distinguishable transition points at small scales – merge together in the thermodynamic limit ($\beta \to 0$).

Intrinsic advantages of the RESE are their intrinsic simplicity and ability to predict dynamical and statistical properties of the emitted light. Although the approximation on which they are based introduces strong limitations, whose shortcomings will be later discussed, their merit is to provide a wealth of information on class B lasers [13] even at the micro- and nanoscale[3]. Surprisingly, in spite of the over 30 years of this model's existence [6], its full potential has so far remained untapped. Our analysis will highlight interesting matches between experimental observations and RESE predictions.

This paper is dedicated to the memory of Prof. F. Tito Arecchi[4], with whom two of us have studied (the second author is a *second generation* student). It is only fitting that a modern discussion of laser threshold should be chosen to remember him, as he was one of the main actors in the development of laser photon statistics and in the dynamical characterization of the properties of devices which have, in the course of his lifetime, unmistakably and irreversibly marked the development of our contemporary society.

Given the length of this contribution, we have tried to organize it in a way as to render it accessible as a resource. The paper is organized in sections with frequent calls from one to the other for ease of cross-referencing, and, as far as possible, independently readable. The Reader can reasonably consult the section of interest without having to delve into all others – at least at a first reading level. Additional material is provided in the Supplementary Information section – also cross-referenced – where we offer useful supplements which would have otherwise further burdened the main paper. All references (sections, figures or equations) to material in the Supplementary section are preceded by a letter S-, followed by the identification.

Apart from the Introduction (Section I), the organization of the paper is the following (by section number):

II. Presentation of the context in which the investigation is placed, with historical and recent references;

III. List and brief analysis of the different threshold definitions which will be examined in this paper. The equivalence among the definitions in the thermodynamic limit is presented in Section III A;

IV. From an experiment conducted on a micro-VCSEL (Section IV A) we extract the information about the following threshold definitions: Gain-Loss Threshold (Section IV B), Photon Statistical Threshold (Section IV C), Fano Threshold (Section IV D), Relaxation Oscillation Threshold (Section IV E), and Largest Fluctuation Threshold (Section IV F). Other threshold indicators are examined in Section IV G. An examination of the dynamical properties of the laser's photon output (Section IV H) is then performed, as a function of pump, through the analysis of the structure of the second-order (zero-delay) autocorrelation (Section IV H 1), of the radiofrequency power spectrum (Section IV H 2) and of the time-delayed second-order autocorrelation (Section IV H 3). A summary of the observations is offered in Section IV I.

---

[2] The designation *thresholdless laser* had been introduced in conference proceedings, in Japanese, cited in [6]. The strictly linear behaviour exists when nonradiative relaxation channels are neglected [9], otherwise a residual distortion in the input-output response persists.

[3] Dynamical class B lasers are characterized by the fast relaxation

of the material polarization and thus describe very well the almost totality of micro- and nanolasers – all based on semiconducting materials.

[4] Originally intended to be part of the Special Issue dedicated to his memory, this contribution arrived too late to be included in the collection.

V. A brief overview of some crucial properties of the REs (thermodynamic limit) is offered in Section V.

VI. This section recalls the definition and the main properties of the RESE model, with a discussion of the changes to the model's solution induced by the introduction of the spontaneous emission contribution (Section VI), including the explicit consideration of the laser's input-output characteristics close to the experimental conditions (Section VI A).

VII. A topological analysis of the RESE is presented, concentrating first on the eigenvalues (stability properties) for a microlaser comparable to the one used in the experiment (Section VII A), then considering, in the general case, the changes in the eigenvalues spanning the whole range from nanolasers to macroscopic devices (Section VII B). The convergence towards the thermodynamic limit (very small values of $\beta$) is studied in Section VII B 1.

VIII. The consequences of the information obtained from the stability analysis of the previous section are analysed, concentrating on the Fano Threshold (Section VIII A) which offers an excellent illustration of the impact of the topology on the statistical properties. Further remarks on experimental dynamical features are added in Section VIII B.

IX. The failure of the RESE to identify a true laser threshold is explained on the basis of the assumption which underpins the introduction of the spontaneous emission. The model requirements for a proper description of the laser threshold are presented (Section IX A), together with an interpretation of the reasons why the RESE is so successful in reproducing most of the observed physical dynamics (Section IX B).

X. Conclusions on the findings of this manuscript, accompanied by some heuristic description of the physics of the laser threshold are offered here. Some possible misconceptions about the presence or absence of a threshold in small lasers, as well practical suggestions for its experimental detection are also offered. Finally, the extension of the main results to other small systems is discussed.

## II. CONTEXT

Shrinking the cavity volume had been a long-standing goal since the achievement of cw operation of semiconductor lasers at room temperature [14–16]. Their small footprint and extremely high energy efficiency, compared to the alternative designs of the time, promised compact sources which could operate with extremely low power expenditure (cf. [17] for a review). However, it became soon evident that the cavity volume reduction was calling into question the recognition of threshold [6, 9], necessary for ensuring the coherence properties of the emitted field. Since then, the definition of laser threshold has become a strongly debated issue which has not only engaged the scientific community [18–24], but even prompted a top journal to issue guidelines for acceptance of manuscripts which claimed lasing at the nanoscale [25]; curiously, one of the best indicators for coherence (second-order zero-delay autocorrelation [26]) was left out of the list [25]. At the end of this manuscript we will partly understand a possible rationale for this choice, although the autocorrelation remains one of the best tools for gaining insight into laser characterization.

The interest behind the scale reduction is twofold: on the one hand enabling applications which range from faster modulation speed for telecommunications [4, 27–33] to high-density optical chips which hold high hopes for lower energy consumption in datacenters [30, 34–38]; on the other hand gaining a clear understanding of the physics of nanodevices to enable controlled applications in the field of fully or partially coherent sources [23, 24], such as illumination [39] or sensing [40].

### 1. New phenomena around threshold

Extensive experimental work conducted in small lasers in the last decades has shown that the transition from incoherent to coherent emission differs from the one observed in macroscopic laser physics: a sudden jump from zero- to full-coherence. Rather, spontaneous squeezing has been, for instance, observed and explained with *ad hoc* modelling [41] together with forecasts for sizeable transient squeezing due to pulsed pumping [42], squeezing with cw pumping [43], sub- and superradiant emission [44], or superthermal emission due to mode competition [45–49]. The influence of physical parameters on the laser threshold are reviewed in [50], whereby the experimentally observed transition from LED-like emission to lasing [51], is explained with the help of RESE [52].

Besides the complexity introduced by the interaction among emitters or modes, however, there is a fundamental, and general, emission regime which renders the evolution towards the laser threshold more complex than the simple jump observed in macroscopic lasers: the appearance of spontaneous photon bursts, as precursors of lasing in a way similar to Amplified Spontaneous Emission [53, 54]. It is an intrinsic consequence of the internal time constants (cavity losses and carrier relaxation rate) and of the properties of noise build-up, ultimately responsible for the transition towards coherent emission, which exists even in the absence of other emission regimes (e.g., squeezing). Spontaneously occurring photon bursts have been observed in microlasers [55, 56] and in smaller devices [53, 57, 58], also giving rise to spontaneously occurring superthermal light [59], explained through the stochastic dynamics of the radiation matter interaction [54, 60, 61].

## 2. What is threshold?

In spite of the complexity of the transition from incoherent to fully coherent emission, of which we have offered a small glimpse, the question of threshold characterization has been pursued in more general terms, with conflicting points of view. A very interesting and physically fundamental definition, named *Quantum Threshold* [62] (**QT**), posits that once *one photon* occupies *in average* the lasing mode – as identified by the cavity mirrors – the transition towards coherent emission takes place, since stimulated emission starts a runaway process, stealing spontaneous photons from the other cavity electromagnetic modes[5]. This definition, has a very fundamental value and, aside from a physically significant factor 2 [9], coincides in principle with the macroscopic laser threshold. The interpretation of QT given in [9, 62] corrects a conceptual mistake [7] where one single photon (not average!) in the lasing mode is deemed to be sufficient for lasing (at $\beta = 1$). The conclusion was based on a dynamical argument which considered only the exponential growth of the lasing field regardless of losses through cavity mirrors. This led to the extrapolation of lasing to *zero-threshold* since in this picture the first photon was sufficient to produce lasing. This picture has a strong intuitive appeal but does fall short on the predictive side when it is extrapolated to predict threshold. The implications of this definition and the physical discussion of indicators suited to identify the quantum threshold [62] are very interesting, but will not be examined any further in this paper, for lack of space.

The thresholdless laser regime ($\beta = 1$) is only the most spectacular feature of the RESE [6]. In general, these equations predict a progressively smoother transition between the emission branches as the cavity volume is reduced (thus increasing $\beta$). The smoothness of the transition renders the threshold determination rather inaccurate, a problem which is ascribed, within this framework, to the practical difficulty of determining the pump value in the absence of a sharp kink (cf. Figs. S-11 and S-12 for examples).

Taking an opposite stand, a quantum-statistical analysis of the laser [8] proved that the concept of threshold, used in [6, 63], is not adequate and results from an extrapolation from the macroscopic regime – where the strict definition which applies in phase transitions still holds to a very satisfactory degree – to devices for which no correct definition is possible. While proposing an extension of a threshold definition to smaller devices, it was recognized that no threshold could be properly given in a finite-size system [8] and that the proposed prolongation had a very limited range of validity. The indisputable self-consistency of the arguments does not, however, solve the practical problem of needing at least an operational threshold definition for applications; as a consequence several approaches were used in an effort to obtain a workable concept.

## 3. A shift of paradigm?

A critical analysis of various approaches is offered in [64] which examines different threshold definitions and finds them in disagreement with each other. It offers a reasonable interpretation of the term *thresholdless* as *lacking threshold behaviour* as opposed to possessing a threshold *at infinitely small pump rates*. However, the lack of consistency in the definitions led the Authors to conclude that "a more detailed theory is needed to define a laser's threshold" … " or we should change our assertion that the onset of laser action in small devices is a critical phenomenon" [64].

It is fair to say that both statements are simultaneously true: 1. lasing in small devices is not a critical phenomenon in the sense of phase transitions; 2. to describe the threshold properly, a theory needs to account for the incoherent **and** coherent fields on both sides of the transition[6].

The traditional, macroscopic laser models [66–69] neglect the spontaneous emission, given its extremely small contribution above threshold and the abruptness of the transition. This approximation is mirrored by the traditional REs (e.g., [70], cf. also Section V). Thus, in spite of the incomplete treatment of the problem, the results are self-consistent in the thermodynamic limit and a threshold can be properly identified. In this sense, [64] is correct in asserting that "only a more detailed theory" can solve the threshold question when considering smaller lasers.

On the other hand, it is also true that the threshold in small-sized devices "is not a critical phenomenon" in the statistical sense (already stated in [8]), thus the transition needs to be defined in a different way. By pointing to [71], Ref. [64] hints at alternative approaches which would replace the traditional quantum-statistical description, whose validity is restricted to the thermodynamic limit (and thereabouts) [8]. A recent model [65] takes up this challenge through the inclusion of coherent and incoherent quantum fields for all pump values. In this way, the laser threshold emerges as a dynamical bifurcation of the coherent quantum field which grows from its below-threshold 0 value. This description is independent of system size and its predictions hold for any laser from $\beta = 1$ to $\beta \to 0$.

While Ref. [65] may represent the conceptual breakthrough in modelling the laser threshold, it is too early to see its full implications. In this work, we will only refer to it as the best current attempt at defining threshold, while concentrating on understanding the apparent

---

contradictions which emerge from other traditional definitions.

#### 4. Understanding the issues

A sound understanding of the laser threshold requires not only the paradigmatic shift outlined above, but also a clear explanation of the reasons why different traditional threshold definitions lead to diverging results. Overstepping the last question would only contribute to increasing the confusion, since in the absence of a clear picture one could always find good reasons for preferring one definition over the other. The rest of this paper is going to tackle this aspect of the challenge.

Through evidence experimentally collected in a microlaser, we address the questions which arise when trying to extract a threshold estimate from the data on the basis of standard definitions. The results provide diverging predictions even in a device where there is a sufficiently well defined transition between two emission branches. The analysis of the topological RESE properties, which have remained an untapped resource for over 30 years [6], sheds a substantial amount of insight into the reasons for the disagreements. Furthermore, it also explains an abundance of dynamical properties of the emitted radiation which hold a strong potential for improving our understanding and exploitation of different micro- and nanolaser properties [22, 53, 56, 58, 72–78].

As common to models, the RESE rests on hypotheses which have their limitations, thus it cannot provide answers which lie outside its realm of application. As clarified in Section IX B, its power resides in the correct description of numerous physical processes which lead to the growth of a coherent field. However, its failure is its inability to identify a true laser threshold (i.e., based on a bifurcation, cf. Sections VI and IX B)).

Finally, it is important to remark that the physics which emerges from the experimental observations is not peculiar to micro- and nanolasers, but, rather, applies to lasers of all sizes. This point will become clear from the RESE analysis conducted at all scales. The physics does not change with laser size: it is the parameter interval which shrinks in macroscopic systems to the point where all phenomena collapse into a sole pump value in the thermodynamic limit. Likewise, the physical threshold does not disappear while shrinking the laser size. It is its representation, evolved over decades of work on macroscopic devices, and the habits acquired in handling definitions which hold in the thermodynamic limit which have provided us with a distorted view of what a laser threshold ought to be. It is **not a sudden change in emission properties** (directionality, coherence, amount of power emitted in the lasing mode), but, as mentioned in Section II 3, the (more or less) gradual **emergence of a coherent field**. Thus, a threshold must exist in all lasers, irrespective of their size, even though we do not immediately recognize it if we apply old stereotypes.

### III. THRESHOLD DEFINITIONS

Over the years, many ways of identifying laser threshold have become common. The reasons for the multiple definitions is often grounded either on the experimental availability of the information[7], or on the appropriateness of a concept to a particular line of investigation. The fact that in macroscopic lasers the different definitions lead to the same result has engrained the idea of their equivalence. However, as we will see by examining some of the main definitions, this is not the case in smaller systems: already at the microscale we encounter discrepancies which lead to confusion about the threshold concept.

By comparing the predictions of the different definitions, which coincide in the thermodynamic limit (or sufficiently large lasers), to those observed in microlasers (Section IV, interpreted in VI) we highlight their discrepancies and try to retrace their origin. The main result is that the differences become apparent only in sufficiently small devices, where the physical processes which lead to the transition from incoherent to coherent emission *unfold* at small scales.

We consider the following threshold definitions to which we assign an acronym for later referral:

1. **BT** (*bifurcation threshold*): lasing begins when the coherent optical field emerges from a bifurcation [79] issued from the change on its stability (at the pump value for which $\Re e\{\lambda\} = 0$, where $\lambda$ represents the multiplicative factor in the temporal evolution $e^{\lambda t}$ of a perturbation).

2. **GLT** (*gain-loss threshold*): lasing begins when the gain experienced by the intracavity field balances *all* losses (refined models explicitly introduce absorption, thus identifying a transparency region where the latter is compensated, while the additional losses – e.g., transmission through the cavity output mirror – are not).

3. **LFT** (*largest fluctuation threshold*): lasing begins when the fluctuations diverge (mathematically) or become extremely large (in practice) as in all phase transitions.

4. **FT** (*Fano threshold*): lasing begins when the Fano factor [8]

$$F = \frac{\langle \sigma_n^2 \rangle}{\langle n \rangle} \tag{1}$$

takes its largest value (infinite in the thermodynamic limit) – $\sigma_n^2$ stands for the variance of the

---

photon number and $n$ for its value, where $\langle \cdot \rangle$ represents the average. A closely related parameter (but less useful for threshold identification) is the Relative Intensity Noise (RIN) defined by

$$RIN = \frac{\langle \sigma_n^2 \rangle}{\langle n \rangle^2} = \frac{F}{\langle n \rangle} \,. \tag{2}$$

5. **PST** (*photon statistical threshold*): The statistical distribution of the photon Probability Density Function (PDF) transforms (abruptly in *infinite-sized* lasers) from Gaussian (below threshold) to Poissonian (above threshold). The transition can be quantified through the zero-delay ($\tau = 0$), second-order autocorrelation

$$g^{(2)}(\tau) = \frac{\langle n(t)n(t-\tau) \rangle}{\langle n(t) \rangle^2} \,, \tag{3}$$

where $t$ represents time in the stream of photons emitted by the laser. The two PDFs are identified through $g^{(2)}(0) = 2$ for Gaussian and $g^{(2)}(0) = 1$ for Poisson statistics.[8]

6. **CT** (*coherence threshold*): lasing appears as soon as a finite value (infinite in the thermodynamic limit) of the coherence time $\tau_c$ or coherence length, $l_c = c \cdot \tau_c$ ($c$ light speed), emerges. This is operatively quantified by the time constant $\tau_c$ with which $g^{(2)}(\tau)$ (exponentially) decays towards $g^{(2)} = 1$.

7. **QT** (*quantum threshold*): the lasing transition takes place when, in average, one spontaneous photon occupies each electromagnetic cavity mode (a similar definition, aside from a factor 2, is given in [9, 62]).

8. **ROT** (*Relaxation Oscillation Threshold*): lasing appears in concomitance with a coupled temporal dynamics between the photon and the carrier number, responsible for the so-called Relaxation Oscillations [80] (imaginary part of the exponent $\lambda$ of the BT definition). This predictor is experimentally accessible through the radiofrequency power spectrum of the temporal photon stream emitted by the laser.

It is important to remark that four of these threshold definitions (LFT, FT, PST, CT) explicitly depend on noise and on the average photon number. However, their functional dependence varies from one quantity to the next, thus hinting to the fact that **if the noise properties change with $\beta$**, then one can expect discrepancies in their predictions.

---

[8] Statistical theory tells that other distributions exist whose values of $g^{(2)}(0)$ match those of either Gaussian or Poisson statistics; thus this identification is not unique. However, since one does not expect other distributions to hold a place in laser physics, the match between PDFs and $g^{(2)}(0)$ values can be considered to be univocal.

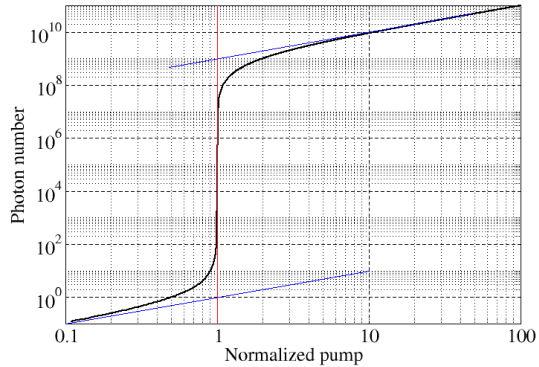

FIG. 1. Illustration of the jump in photon number in the lasing mode at threshold for a large laser (solid line, black online, $\beta = 10^{-9}$), traced with the help of RESE (cf. Section VI). The construction with straight lines extrapolates the response to that of an infinitely large laser (thermodynamic limit) where the upper straight line should move to $\infty$ and is drawn here to illustrate the construction procedure. The two (blue online), parallel straight lines represent the growth in the photon number due to spontaneous (lower branch) and stimulated emission (upper branch) and asymptotically align with the (large but) finite-size laser response (solid line). The vertical line (red online) represents the discontinuous jump from the lower to the upper emission branch (incoherent to coherent lasing transition) and passes through the mid-point of the connection between the two branches (solid line), which coincides with the value of the normalized pump equal to 1. The intersection between the vertical line and the lower emission branch represents the average photon number at the pump value for which the transition occurs (threshold). It conceptually coincides with the QT and corresponds to the announced value of the average photon number: 1. The mid-point in the steep portion of the curve – through which the (red) vertical line is traced – matches the GLT, which then coincides with the QT.

Finally, as summarized in Table I some of the previous definitions have a purely conceptual value or match theoretically/numerically accessible quantities, while others correspond to experimentally measurable variables.

## A. Equivalence among definitions (thermodynamic limit)

Some simple physical considerations illustrate the reason why – for very large lasers – the previous threshold definitions can all be considered equivalent. In the rest of the paper, we will examine experimental results which tell a different story and interpret them with the help of the RESE model.

The equivalence among definitions, in the thermodynamic limit, stems from the following considerations. We start from the primary, BT, definition, which is based on the analysis of the temporal evolution of an arbitrary fluctuation $\vec{\xi}$ imposed on a variable $\vec{x}$ (in a space with arbitrary dimension $M$) around its equilibrium point $\vec{x}_s$ (of unknown stability) determined by (at least) one parameter – the pump in the laser case. Its evolution, at least at short times, can be cast in the form

$$\vec{x}(t) = \vec{x}_s + \vec{\xi}e^{\lambda_j t}, \qquad (4)$$

with $\lambda_j$ (complex) coefficient of the evolution in the $j$-th direction (with $1 \leq j \leq M$), already used in the definition of **BT** (for a single dimension). $\Re e\{\lambda_j\} < 0$ implies relaxation (along the $j$ direction) towards $\vec{x}_s$, while $\Re e\{\lambda_j\} > 0$ corresponds to a divergence away from it, towards a new state (if it exists). The bifurcation occurs when $\Re e\{\lambda_j\} = 0$: in this case, an arbitrary perturbation $\vec{\xi}$ leads to a new equilibrium condition, since no temporal relaxation can take place (indifferent equilibrium). Thus, an accumulation of successive fluctuations $\vec{\xi}$ at this control parameter value (again, pump for the laser) can lead the system very far from $\vec{x}_s$.

TABLE I. Nature of the different threshold definitions. The left column regroups predominantly conceptual definitions, while on the right hand side appear more practical techniques.

| 1. | **BT**: Conceptual definition. Information difficult to access experimentally. **Primary and sole** proper definition of laser threshold. | 4. | FT: Operationally important. Quantitative information accessible from measurements, even though its quality degrades rapidly when the lasing volume shrinks (Section IV D). |
|---|---|---|---|
| 2. | GLT: Conceptually relevant. Information accessible from input-output response curves (cf. Sections IV B, S-4 A) or through extrapolation (approximate information). Noise-independent. | 5. | PST: Operationally important (Section IV C), especially since it can be directly retrieved through photon counting techniques. Particularly suited to measurements with very low photon flux values. |
| 3. | LFT: Conceptually important. Fluctuations can be measured (Section IV F), but the quantitative information is best retrieved through the FT. | 6. | CT: Operationally important. As PST, it can be directly obtained from photon counting techniques. |
| 7. | QT: Conceptually important, but somewhat controversial [8, 81]. Its definition is not entirely rigorous, although it represents an excellent illustration of the requirements for a transition between the two emission states. | 8. | ROT: Operationally important (Section IV E), although its definition does not necessarily coincide with the BT threshold (cf. Section V for an illustration in the thermodynamic limit). |

This matches the statistical physics-based statement that at the phase transition fluctuations diverge: any amount of deviation from $\vec{x}_s$ is compatible with the state of the system. In practice, in a large, finite-size system the fluctuations will possibly be very large, but never infinite, but this does not detract from the quality of the analogy. However, this proves the equivalence between BT and LFT; since FT is nothing else than a quantitative translation of the LFT, *then the three definitions are equivalent.* It is important to reiterate that the bifurcation-based definition is the most general one, since it does not depend on system size and thus – unlike the statistical definition – applies to any kind of laser.

Conceptually, the GLT, based on the exact equilibrium between amplification and attenuation, corresponds to the pump rate for which a coherent field is no longer attenuated and can start to grow (CT). Since this occurrence can take place only at the bifurcation point (indifferent equilibrium for any value imposed for the photon number), then the two threshold definitions are equivalent. Since when the PDF becomes Poissonian, then the autocorrelation $g^{(2)}(\tau) = 1$ for any value of $\tau$ (in the thermodynamic limit), *then it follows that PST and CT are equivalent,* **thus establishing the equivalence of all these thresholds**.

The equivalence of the QT definition with the previous ones rests entirely onto the notion of thermodynamic limit (infinite system size). This can be illustrated with the help of the schematics shown in Fig. 1a, which depicts the number of photons present in the lasing mode (intracavity) as a function of pump. The solid line is traced with the help of the steady state of the RESE, which will be discussed in Section VI, for $\beta = 10^{-9}$, giving a good approximation of the thermodynamic limit. Nonetheless, the black curve is continuous and thus departs from the strict limit $\beta = 0$. Reconstructing the equivalent sketch for the thermodynamic limit[9] enables the visualization of the sudden step, shown by the discontinuity between the jump (red line) and the lower blue line, which represents the spontaneous emission: the red and lower blue line meet at an average photon number $\langle n \rangle = 1$ for the value of normalized pump 1 (representing threshold). This picture illustrates the equivalence between the statistical threshold, in the thermodynamic limit, and the QT, for which[10] $\langle n \rangle = 1$. Since the LFT corresponds to the statistical definition of phase transition (the vertical red line in Fig. 1), we can consider the equivalence between the QT and all the previously examined threshold proven.

Finally, we are left with the examination of the ROT.

_______

[9] The two blue diagonal lines extrapolate the asymptotic behaviour of the spontaneous, lower, and stimulated, upper, branches to enable the determination of the intersection with the vertical jump (red line), chosen to pass through the point at mid-height of the steep-sloped connection. In reality, in the true thermodynamic limit the upper branch should be at infinity, since the photon number at the jump is $\sim \beta^{-1}$.

[10] We are here using the definition for which one photon in average is present in the lasing mode at QT, at variance with [9, 62].

The latter results from a physical approximation (in the thermodynamic limit) which rests on a strong plausibility argument: once a macroscopic coherent field exists inside the cavity, its interplay with the population inversion (represented by the bilinear term $nN$ in eqs. (5,6) – cf. Section V) gives rise to oscillations between the two physical quantities (which vary in time with a quadrature shift, where the population leads the photon number). In reality, we will show in Section V that the ROT does not coincide with the previous definitions, but occurs for a pump value that is slightly larger: $\approx \frac{\gamma}{4\Gamma_c}$ [82], where $\gamma$ and $\Gamma_c$ are the relaxation rates for the population inversion and intracavity photon number, respectively (cf. eqs. (5,6)). In practice, the distance between ROT and the other thresholds is small enough to be experimentally indistinguishable in large lasers.

The advantage of the ROT is its experimental accessibility, at least in micro- and macrolasers, since it can be obtained from the radiofrequency (rf) power spectrum, which signals the emergence of the Relaxation Oscillation (RO) frequency [80]. Thus, even though the ROT is not strictly equivalent to the other thresholds, it is sufficiently close for most large lasers to coincide within the experimental precision.

## IV. EXPERIMENTAL RESULTS

We now examine the various threshold definitions in the context of the experimental observations obtained from a micro-VCSEL characterized by an estimated fraction of spontaneous emission in the range ($1.8 \times 10^{-4} \lesssim \beta \lesssim 3 \times 10^{-4}$) (cf. Supplementary Information in [55]), i.e., a mesoscale device [83]. After a brief description of the experimental set-up and of the measurement techniques, different subsections examine the threshold information extracted from the measurements.

### A. Experimental set-up

The choice of a commercial device offers the advantage of reliability and durability in time [84], even though the construction parameters have to be indirectly inferred due to industrial secrecy. The measurements are taken on an electrically pumped, small-diameter ($d = 6\mu m$) VCSEL (Thorlabs VCSEL-980) with nominal emission wavelength $\lambda = 980nm$. The VCSEL is mounted on a Thorlabs TCLDM9 module which permits a temperature stabilization at $T = (25.0 \pm 0.1)^\circ C$, thanks to home-built apparatus driving the Peltier element. The pump current is provided by a stabilized current source (Thorlabs LDC200VCSEL, $1\mu A$ resolution, $\pm 20\mu A$ accuracy), a crucial feature – together with temperature stabilization – for the accurate investigation of threshold properties since high stability is needed to distinguish the different dynamical regimes of laser operation. Optical reinjection is avoided through the use of a Faraday isolator (QIOPTIQ 8450-103- 600-4-FI-980-SC) and the light is detected by a fast, large bandwidth ($9.5GHz$) amplified photodetector (Thorlabs PDA8GS) coupled to a multimode $60\mu m$ optical fibre. The data are sampled by a LeCroy Wave Master 8600A digital oscilloscope ($6GHz$ analog bandwidth, $0.1ns$ sampling time) which stores up to $5 \times 10^6$ samples per trace. Additional details on the basic set-up can be found in [55] and its Supplementary information file.

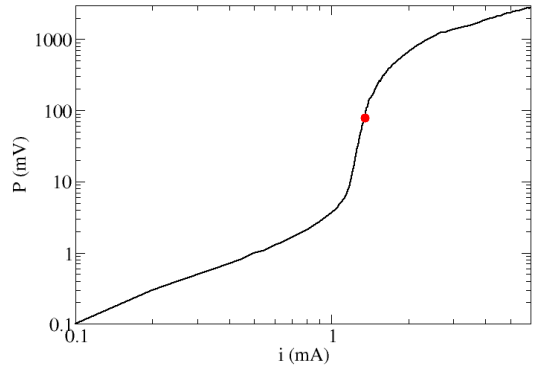

FIG. 2. Laser input-output curve obtained from the VCSEL-980 (Thorlabs). $P$ represents the laser power (measured in mV by a preamplified UDT-455 detector), while $i$ stands for the pump current. The dot (red online) marks the position of the GLT (at $i \approx 1.36mA$).

The measurements are taken fixing the current at the desired value, keeping all the controls active and acoustic protections in place (which also prevent air currents), and taking in rapid succession 10 sequences of $10^5$ points at a sampling rate $0.1ns$. The data are then stored onto a computer for further processing. In the following, the data are treated to obtain averages from the 10 sequences, from which error bars can be extracted. Background noise measurements are also taken and their average value is subtracted from the data prior to processing.

Measurements taken on other samples of Thorlabs VCSEL-980 provide results which are compatible with those reported here with pump current values which may vary up to $\approx 20\%$, depending on the device. The deviation from one device to the next is not negligible, but the reproducibility is good, proving the soundness of the choice of commercial devices.

Finally, it is important to remark that the input-output measurements are taken fixing the pump current and taking an average reading of the power impinging onto the detector. Performing an automatic scan of the current would lead to a displacement of the threshold [85], thus introducing a bias in the determination of the threshold position.

### B. Gain-Loss Threshold – GLT

Repeating the construction used in Fig. 1, with the experimental branches below and above threshold, we compute geometric mean value [8] of the laser power, thus de-termining the current value for the GLT ($i \approx 1.36\,mA$). The point is identified by the circle (red online) superposed onto the laser characteristic curve (Fig. 2). The asymmetry in the lower and upper curvature connecting the two branches makes the point appear somewhat displaced towards the upper branch; however, the result is consistent with the asymptotic growth of the laser power.

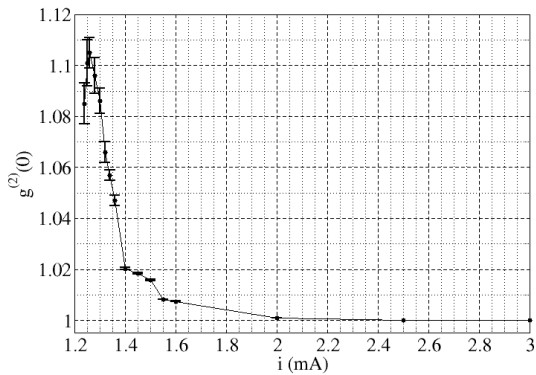

FIG. 3. Second-order, zero-delay autocorrelation obtained from the experimental data as a function of pump current. The error bars are computed on the repetitions of the auto-correlation taken from 10 files of $10^5$ datapoints each. When not visible, the error bars are smaller than the dot size and become negligibly small at $\approx 2\,mA$. In this and the following figures the line connecting the experimental points is added to guide the eye.

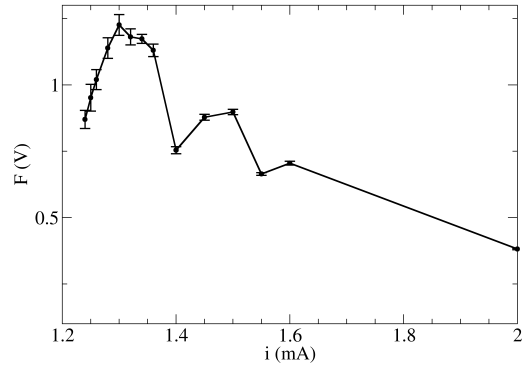

FIG. 4. Fano factor computed from the fluctuations. The 10 datasets per point enable the computation of the error bar, as in the previous figures.

### C. Photon Statistical Threshold – PST

The computation of $g^{(2)}(0)$ (eq. (3)), for $\tau = 0$ from all the experimental data provides the functional dependence which appears in Fig 3. The entirety of the curve will be exploited later (cf., e.g., Sections IV F, IV H and IV I). For the moment, we are interested in searching for the $g^{(2)}(0) = 1$ value, which appears[11], within error bars, at $i \approx 3\,mA$ (the graphical resolution may suggest an earlier value, but a close look at the data belies this conclusion). The PST is very far from the GLT!

Following [8], if we choose the GLT as reference the PST is reached at a normalized pump value $\frac{i_{PST}}{i_{GLT}} \approx 3.3$. In other words, coherence is obtained at a pump 3.3 times larger than what expected[12].

---

[11] In reality, more accurate measurements performed with photon counting techniques – which suffer from a much smaller bandwidth limitation compared to the linear detection used in these measurements – show for this device that the Poissonian statistics is only reached at $i \approx 4.5\,mA$ [59]. For the purpose of this discussion the estimate coming from Fig. 3 is an acceptable approximation.

[12] It is interesting to notice that according to the manufacturer, the threshold current is in the range $2.2\,mA \leq i_{th} \leq 3.0\,mA$

Since the CT is reached at the same time as the PST, the two coincide and do not need an independent measurement.

### D. Fano Threshold – FT

Computing the average fluctuations from the datasets and the corresponding averages (eq. (1)) we obtain the Fano function displayed in Fig. 4. Aside from the dips at $i = 1.40\,mA$ and $i = 1.55\,mA$, on which we comment later (cf. Section IV H 1), the most striking feature of this function is the absence of the expected sharp peak, which is supposed to mark the FT. At least three points could be considered as belonging to a broad maximum $1.30\,mA \leq i_{max} \leq 1.34\,mA$, just restricting the analysis to one standard deviation (plotted error bars). This is in clear contradiction with the expectation of a clear transition marked by a cusp [8], typical signature of threshold as a phase transition.

### E. Relaxation Oscillation Threshold – ROT

A clear resonance peak in the radiofrequency power spectrum appears at $i = 1.40\,mA$, slightly above the GLT, suggesting that the ROT lies close to the latter, but slightly above (by approximately 3%). A lack of coincidence between these two threshold definitions agrees even with the RE model written in the thermodynamic limit (Section V), even though here the quantitative disagreement is much larger than what the REs forecast. Comparison with the RESE model is more complex (cf.

---

($1.6 \leq \frac{i_{th}}{i_{GLT}} \leq 2.2$). Comparing to Fig. 3 we see that for practical purposes the manufacturer's value is in reasonably close agreement with our estimate of the PST, but remains far from the other threshold definitions.

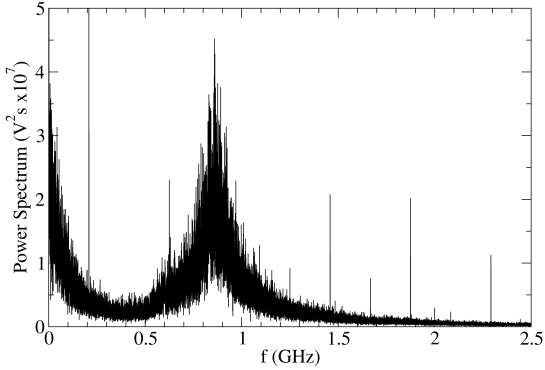

FIG. 5. Radiofrequency power spectrum at $i = 1.40mA$.

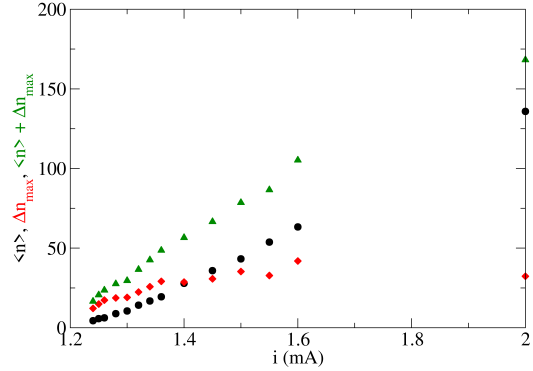

FIG. 7. Largest absolute photon number $n_{max}$ (triangles, green online) as a function of pump current. The largest fluctuation from the average ($\Delta n_{max} = n_{max} - \langle n \rangle$) is represented by the diamonds (red online) while the average ($\langle n \rangle$) by the circles (black online).

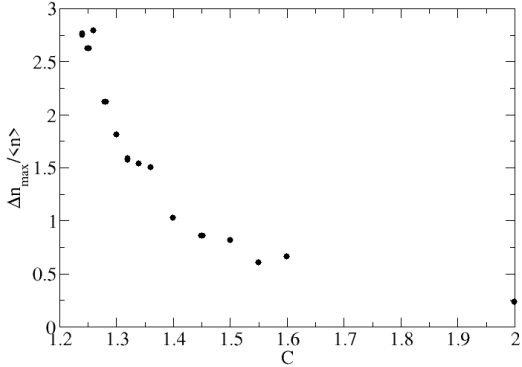

FIG. 6. Largest relative photon number $\Delta n_{max} = \frac{n_{max} - \langle n \rangle}{\langle n \rangle}$ as a function of pump current.

Section S-4 F), but the lack of coincidence in pump value persists even there. We will later see how this information fits into the larger picture of the rf spectrum of the laser output in the complex transition between incoherent and coherent emission (Section IV G).

The interest of the ROT lies in the fact that it can be easily identified in experiments and that a related technique has been shown to enhance its identification in microlasers (and is expected to also efficiently perform for nanolasers) [86].

## F. Largest Fluctuations Threshold − LFT

The concept of largest fluctuation is not easily implementable in an experimental context, as the observations are of necessity limited in number: it is never possible to know to what extent the sample that has been gathered includes the largest fluctuation. Nonetheless, given a sufficiently large number of datapoints ($10^6$ for each pump value in our measurements) it is possible to search for the maximum fluctuation (largest deviation) and observe the trend.

Fig. 6 shows the largest photon number divided by its average, as a measure of the largest relative fluctuation observed as a function of pump. The trend, not unexpectedly, overall resembles that of the second order autocorrelation (Fig. 3), with the largest fluctuation observed around the minimum pump value used for the measurements. Looking at the data, it is obvious that the largest fluctuation can only occur for $n > \langle n \rangle$ (a fluctuation below $\langle n \rangle$ being limited by $n \geq 0$), thus, for consistency, we plot $n_{max}$ throughout. The conclusion emerging from the graph is that the largest relative fluctuation occurs at the lowest pump values. The physical origin of this occurrence is related to the temporal dynamics (cf. Section IV H 1) and exits the realm of all currently existing models [54, 56], with the exception of fully stochastic laser simulations [60, 61, 87, 88] and of one *ad hoc* modification of the RESE [61].

Looking at the absolute largest photon number, as a function of pump current, we observe a steady growth with non-constant rates (Fig. 7, triangles, green online). This somewhat surprising occurrence is best understood by decomposing the maximum photon number into $n_{max} = \langle n \rangle + \Delta n_{max}$, where the average photon number growth is represented by circles (black online) and the largest fluctuation from the average by diamonds (red online). The picture is now clearer, as we observe a rapid growth of the largest fluctuation, whose amplitude is equalled, then surpassed by the growth of the average photon number at $i = 1.4mA$. Thus, the steady growth of the maximum photon number actually consists of two separate trends, where the fluctuation dominates initially, then steadily reduces its influence, as observed in Fig. 6. The conclusion reached previously is therefore confirmed by this picture.

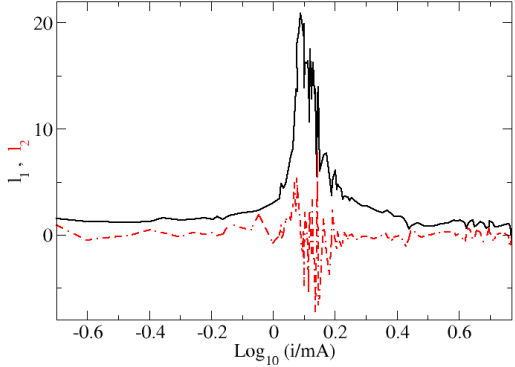

FIG. 8. Experimental first-order, $l_1$ (solid line, black online), and second order, $l_2$ (double-dash-dotted line, red online), logarithmic derivatives of the laser power as a function of the pump current, computed according to the numerical scheme explained in the text. The variables are normalized to $1 mV$ for $P$ and $1 mA$ for $i$ (to compute the logarithm).

### G. Other threshold indicators

Even though not part of the listed threshold definitions, we consider another intriguing indicator deemed capable of determining threshold in nanolasers. It is based on the features which characterize phase transitions in statistical mechanics and has been proposed in a well-cited paper [19].

Two complementary quantities are proposed which, resting on the differential variations of the photon flux as a function of pump, exploit the sensitivity of the derivative to minor changes which go unnoticed in the functional dependence. Their accrued, differential sensitivity highlights the presence of details which define a threshold. Using the definitions introduced in [19] adapted to the experimental datasets (as explained in Section S-1 B), we plot the logarithmic derivatives $l_1$ and $l_2$ (Fig. 8).

The second derivative $l_2$ (double-dash-dotted line, red online) is to be compared to the curves computed in [19] from the rate equation model and shown there in Fig. 3 (bottom panel, [19]). In the theoretical paper [19], excellent threshold predictions follow from the determination of the maximum of $l_2$ for a microlaser (shown there for $\beta = 10^{-3}$ – our results in a lower $\beta$ device should be even better). Instead, in place of a bell-shaped curve we obtain a functional dependence which oscillates around 0 (Fig. 8). This can hardly be associated with a clear threshold definition.

The problem clearly comes from the predictor's dependence on the second derivative of the laser's experimental response. Every derivative amplifies the measurement noise, which ends up dominating: while in the first derivative (solid line, black online) there persists a trace of a noisy functional dependence, the second one does not resemble in any form the expected predictor [19].

Some filtering could be used to smooth the experimental data and the ensuing functional forms, (cf., e.g.,

Figs. S-3, S-4 in Section S-1 B), but it is very unlikely to improve the results to the point of obtaining reliable quantitative predictions on the threshold pump. Keeping in mind that microlaser measurements, thanks to the larger photon flux, are far less noisy than those obtained from nanolasers, one must *a fortiori* expect poorer predictions at the nanoscale.

We can therefore conclude that, although the phase-transition-based definitions proposed in [19] are sound and grounded in fundamental physical features, they are not usable to provide information on the laser threshold from experimental data. This conclusion makes sense when considered in the context of small systems: phase transitions, and their statistical indicators, are observed in infinite-sized systems [8], thus their indicators are suitable to those scales. Small systems are characterized by a high degree of noise (in spite of their overall higher dynamical stability – cf. Section VII B), thus predictors which amplify noise cannot be suited at these scales.

### H. Analysis of the transition dynamics

The application of the different thresholds to the experimental data obtained from the micro-VCSEL shows not only the inherent difficulties in using the various definitions, but also that discrepancies can be enormous (cf. Section IV I). The physical origin of this disagreement resides in the existence of a complex dynamics which develops in the pump region where the electromagnetic field gradually acquires coherence. This region is entirely unknown in the physical picture based on macroscopic lasers (cf. Section V) and is intimately related with the finite-size nature of the problem. Thus, it escapes all description based on phase transitions and the thermodynamic limit of the laser [8, 89–92].

Some of the results shown below have been already published, but the following is a detailed and comprehensive review of the ensemble, which allows for the construction of a more complete picture and of a deeper understanding of the lack of equivalence of the different definitions.

#### 1. Structure of $g^{(2)}(0)$

First of all, the evolution of the zero-delay second order autocorrelation clearly deviates (Fig. 3) from the standard jump from thermal ($g^{(2)}(0) = 2$) to Poissonian ($g^{(2)}(0) = 1$) value, typical of very large lasers [26] (cf. also Fig. S-15, for $\beta = 10^{-8}$): the shape of the transition is gradual and, most important of all, possesses a structure so far not found in analytical models[13].

Fig. 9 shows an ensemble of sampled dynamics at various important points of the autocorrelation function. A progression through large (relative to the background) and for a good part sharp photon bursts (panel (a)) gradually gives rise to a growth in the continuous part of the emission (panels (b) and (c)) eventually leading to a dynamics distinguished by a much more stable time-independent component of emission with superposed irregular oscillations (panel (d)). While the first three panels match the steeper descent in the autocorrelation function from its peak (Fig. 3), the fourth signals the onset of the first "plateau". Panel (e) shows a much more regular dynamics, with noisy, but less disordered oscillations at a frequency which has clearly increased. This dynamics coincides with the gradual drop towards the Poissonian emission, whose dynamical precursor is displayed in panel (f) (the Poissonian limit is reached at much larger pump values, thus noise is still rather sizeable here). Notice that the vertical scale changes from panel to panel to best display the dynamics and that bandpass filtering has been applied (cf. caption of Fig. 3) to remove the low-frequency components of electrical noise that most disturb the picture. All the subsequent figures have been computed on data traces on which filtering has been applied.

---

[13] Computing $g^{(2)}(0)$ with an entirely stochastic numerical scheme [60] qualitatively well reproduces [55] the main features observed in the experiment.

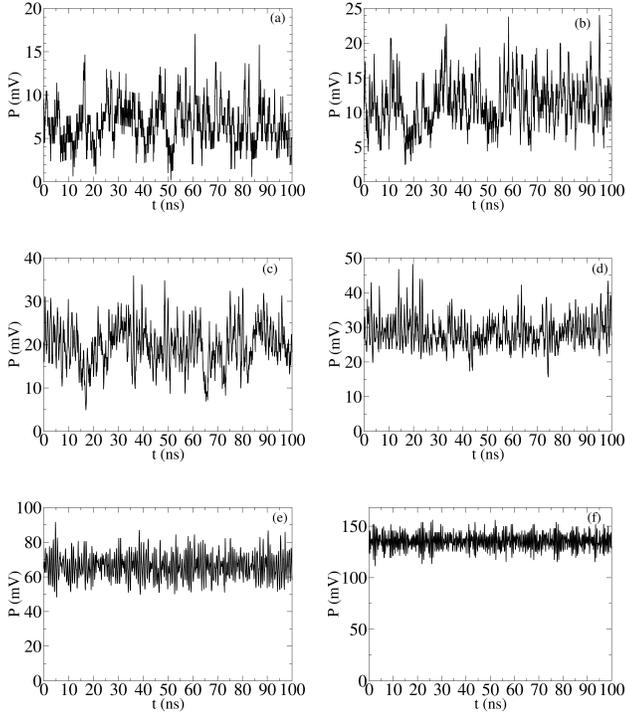

FIG. 9. Samples of temporal dynamics of the laser output for selected current values ($i$ in $mA$): 1.26 (a), 1.30 (b), 1.36 (c), 1.40 (d), 1.60 (e), 2.0 (f). Bandpass filtering of noise between $0.5MHz$ and $5MHz$ has been used to remove those components of background noise which most disturb the traces.

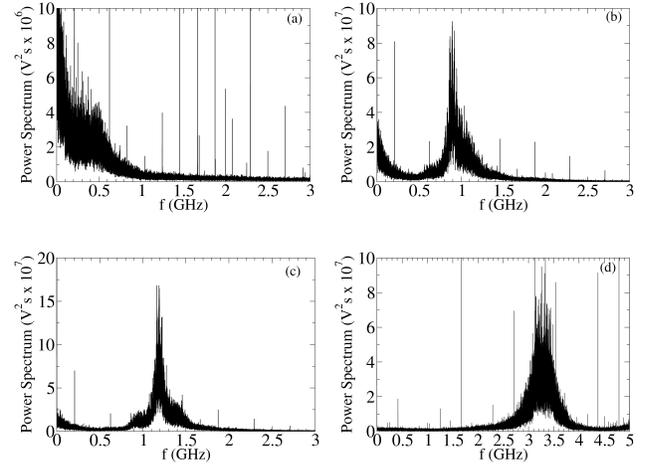

FIG. 10. Average radiofrequency power spectra of the laser dynamics at $i =$: (a) $1.26mA$, (b) $1.45mA$, (c) $1.60mA$, (d) $2.0mA$. Notice the changing vertical scales in the different panels and in the horizontal scale for (d). The sharp lines correspond to sampling noise from the oscilloscope.

The more noticeable changes in the shape of the autocorrelation are therefore linked to the kind of temporal photon emission. The more irregular, larger fluctu-ation bursts [56] are recognizable by the larger values[14] of $g^{(2)}(0)$, with a very steep descent. As soon as the signal acquires a semblance of (noisy) oscillation, the autocorrelation values settle down and show only a smaller decrease, until a more regular oscillation (panel e) sets in to slowly decay towards Poissonian statistics. This last transition is accompanied by the growth of a substantial continuous component of the emission with relatively smaller oscillations.

A trace of these changes in temporal dynamics, reflected in the shape of the autocorrelation (Fig. 3), can also be found in the unexpected dips in the functional dependence of the Fano factor (Fig. 4). This is not surprising since both indicators rely on the variance of the photon number – albeit with a somewhat different functional dependence –, thus it is not unusual to find its footprint in both.

### 2. Frequency power spectrum

Further support to the previous remarks comes from the rf power spectra (Fig. 10). Panel (a) shows the spectrum for the dynamical photon output displayed in Fig. 9a: aside from the normal large low frequency component, no particular time-scale emerges for the temporal signal (a hint of a "resonance" around $f \approx 0.5GHz$ can be appreciated). The relatively flat spectral features suggest a lack of a strongly characteristic timescale for the emission of the photon bursts displayed in Fig. 9a, while the rapid cutoff with frequency shows that the phenomenon cannot repeat at arbitrarily fast speeds.

---

[14] Remember that the absolute values of $g^{(2)}(0)$ in the figure are small due to the filtering action of the data acquisition chain. The important point is the relative behaviour of the function.

These features remain substantially unchanged, with only a gradual emergence of a preferential frequency component which matches a repetition rate, rather than an oscillation frequency[15]. The value of the frequency component which marks the shoulder of the broad spectrum increases from $\approx 0.5\,GHz$ to $\approx 0.8\,GHz$ at $i = 1.36\,mA$ (not shown).

When $i = 1.40\,mA$ a sharp change takes place (later recognizable in the delayed autocorrelation function Fig. 11d), which is now distinguished by a clear resonance frequency in the rf power spectrum at $i = 1.40\,mA$ (Fig. 5). Comparison with the temporal signal confirms its oscillatory nature which progresses with a sharpening of the resonance at $i = 1.45\,mA$ (Fig. 10b). This culminates into a much stronger spectral component ($i = 1.60\,mA$, Fig. 10c, matching the temporal trace of Fig. 9e), with slightly asymmetric shoulders. The stronger and sharper peak suggests the presence of a strong coupling between the two variables (photon and carrier numbers) with a robust oscillation propensity. From this point on, the oscillations strength is reduced and the feature broadens (illustrated in Fig. 10d, corresponding to the signal of Fig. 9f), to converge towards a temporal trace with a very small amount of noise ($i \geq 3.0\,mA$).

### 3. Time-delayed autocorrelation

An even clearer picture emerges when analysing the results of the time-delayed autocorrelation in the context of the previous considerations (Fig. 11). Panel (a) shows a central autocorrelation peak with half-width $\tau_p \approx 0.25\,ns$ and negligible correlation revivals[16] (an extremely faint trace is visible at $\approx 2\,ns$, matching the hint of a spectral feature at $f = 0.5\,GHz$ in Fig. 10a). The lack of revivals proves the mutual independence of the photon bursts observed in the temporal trace (Fig. 9a) and confirms the interpretation of the broad spectral feature (Fig. 10a).

Increasing the current to $i = 1.30\,mA$ (Fig. 11b) broadens somewhat the central peak ($\tau_p \lesssim 0.4\,ns$) and introduces a preferential recurrence time for the bursts at $\tau_b \approx 1.75\,ns$; a second lateral peak can be barely distinguished from the background. It is important to notice that $g^{(2)}(\tau) > 1$ in the whole displayed interval, proving the absence of anticorrelation between photon bursts, thus the absence of a structured temporal evolution. The small degree of regularity which starts to emerge is not sufficient to give rise to true oscillations.

At $i = 1.36\,mA$ (Fig. 11c), the current value for which we estimate the position of the GLT (Section IV B), the autocorrelation acquires a clearly defined structure: in addition to the sharp central peak (half-width $\tau_p \approx 0.3\,ns$) a clear correlated structure appears with bursts occurring preferentially at $\tau_b \approx 1.2\,ns$ and multiples. Four revivals are clearly visible and a couple more can be barely distinguished on the background. We remark, however, that $g^{(2)}(\tau) > 1$ for all values of the delay $\tau$, confirming the lack of anticorrelation, i.e., sinusoidal-like oscillations. We can therefore conclude from the experimental evidence that at this pump value the dynamics is still dominated by photon bursts, which appear preferentially with a periodicity $\tau_b$, matched by a peak which emerges in the power spectrum at $\tau_b^{-1}$ (not shown in Fig. 10). The "regularization" of the emission, albeit without continuity in its dynamics, is a feature which escapes an analysis of the temporal representation, where it would be impossible to draw conclusions on the difference between the traces of Figs. 9b,c.

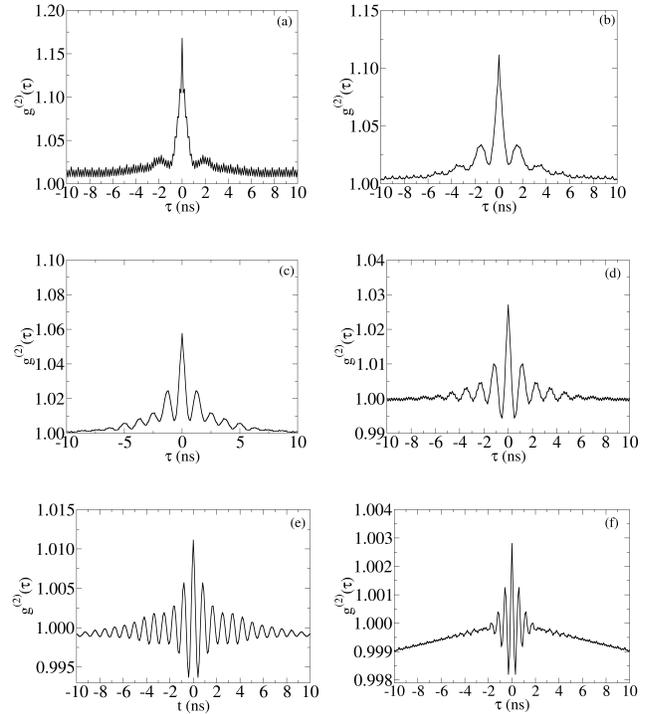

FIG. 11. Average delayed second-order autocorrelation computed on ten temporal time traces ($10^5$ points per trace) at different pump current values ($i$ in $mA$): 1.26 (a), 1.30 (b), 1.36 (c), 1.40 (d), 1.60 (e), 2.0 (f). Notice the slight anomaly in the last two autocorrelation pictures, where the baseline is below 1. In these two traces, covering a very small correlation interval in the vertical scale, the autocorrelation displays the consequences of the bandpass filtering that we have applied to the raw data. The effect is sufficiently small as to not affect the general reliability of the extracted information.

A strong change in the autocorrelation emerges from Fig. 11d ($i = 1.40\,mA$) which corresponds to the signal of Fig. 9d and to the rf power spectrum of Fig. 5; this is the value of pump current to which we have attributed the ROT in Section IV E. The behaviour of $g^{(2)}(\tau)$ clearly shows the appearance of an anticorrelation at $\tau \gtrsim 0.5\,ns$ proving the appearance of an oscillation, whence the as-signment of the ROT. From now on, all autocorrelation functions periodically go below 1, not because of quantum interference effects, but simply because a periodic oscillation sets in through the coupling of photon and carrier numbers.

---

[15] The time-delayed autocorrelation, Fig. 11b,c proves that these are not oscillations.

[16] The small oscillations which appear on the trace are due to the oscilloscope's high-frequency sampling noise.

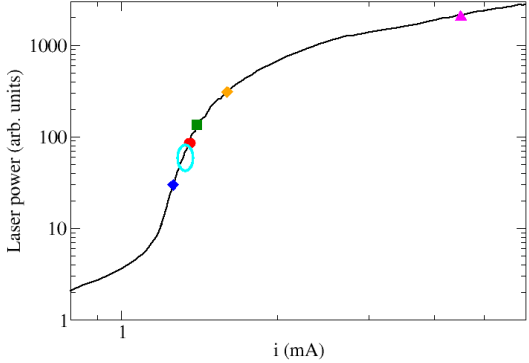

FIG. 12. Experimental laser input-output characteristics with superimposed threshold definitions: FT (open line – cyan online), GLT (circle – red online), ROT (square – green online) and PST (triangle – magenta online) which coincides with the CT definition. The lower diamond (blue online) marks the position for superthermal photon statistics, while the upper diamond (orange online) identifies the pump current value for which the strongest coupled photon-carrier oscillations appear.

The notable change which appears when further increasing the pump current is the strong increase in correlation at $i = 1.60 mA$ (Fig. 11e): oscillation revivals appear in the whole $\tau$-interval displayed (at least 11 oscillations). Thus, there is a strong correlation in the signal, with a larger and narrower peak in the rf power spectrum (Fig. 10c). The correlation diminishes when further increasing the current (Fig. 11f), confirming the information of the power spectrum (Fig. 10d): the photon emission progresses towards the Poisson statistics, where the fluctuations are random and independent of one another.

### I. Summary on threshold definitions

This complex transition scenario has never been fully observed in macroscopic lasers. A partial picture was obtained in solid-state microlasers built as surrogates for semiconductor-based nanolasers [93, 94], but at the time the attention was focussed on obtaining scaling laws which could reproduce the features of nanolasers at the microscale [95]. In spite of the interesting dynamical arguments which were proposed at the time, more recent work on photon statistics [59] has shown that those extrapolations do not hold and that the devices used in [93, 94] still possess a number of features close to those of macroscopic lasers. Nonetheless, the observation in solid-state microlasers of part of the dynamics reported here proves that a gradual evolution exists when progress-ing from the macroscopic to the nanoscopic scale.

The scenario that we have obtained in the previous sections can be summarized by placing the different thresholds on top of the laser response curve (Fig. 12): the GLT is marked by a circle (red online), the FT – due to its poor precision – by the open loop (cyan online), the ROT is marked by a square (green online) and the PST (and, thus, CT) by a triangle (magenta online). Notice the spread, which becomes very considerable when comparing the PST to all other thresholds. The remarkable point is that all these thresholds superpose in the thermodynamic limit (at $C = 1$, Sections VII B, VIII, S-4 C and S-4 F), rendering the definitions equivalent to the primary one (BT).

Fig. 12 shows two additional points (diamonds) which highlight dynamical features missed by the threshold definitions but are relevant for the physics of the transition between incoherent and coherent emission. The leftmost one (blue online) corresponds to the maximum of the zero-delay autocorrelation ($g^{(2)}(0)$, Fig. 3). While not identifiable with a precise threshold feature, this point – corresponding to superthermal statistics [59] – nonetheless stands for an important property of the emitted light, where photons are generated in tighter bunches than those of thermal (or *chaotic*) light. The transition through a superthermal statistics points to dynamical features in the build-up of the coherent field [54, 56] which have not traditionally been addressed in the context of the laser threshold. Yet, they become visible in a mesoscale device, while their presence has not, so far, been identified at the macroscopic scale.

The second diamond (rightmost) marks the position at which the strongest peak in the rf spectrum is observed (Fig. 10c). As already mentioned, it signals a very strong correlation in the coupled photon-carrier oscillations (Section IV H 3). This dynamical observation finds a clear and in principle unexpected explanation in the RESE model (Section VII A).

## V. MACROSCOPIC LASERS

The single-mode theory of the laser has been constructed over the past decades following different approaches (e.g. [66, 96, 97]), which eventually lead to the so-called Maxwell-Bloch form [98] originally proposed in [67] (cf. also [69] for an overview). Based on experimental realizations of the 1960s and 1970s, these models refer to lasers constituted of a very large active volume – and of an even larger cavity – and therefore do not explicitly consider its finite extent. The models therefore consider lasers of ideally infinite size, thus taking the so-called *thermodynamic limit*. In order to ease the analysis and compare more readily to both experiments and RESE, we summarize here the main, well-known properties of these models.

Concentrating on class B laser devices [13], based on active materials where the medium's polarization relaxes very fast and therefore can be adiabatically eliminated[17], the Maxwell-Bloch equations reduce to the standard REs (e.g. [70]).

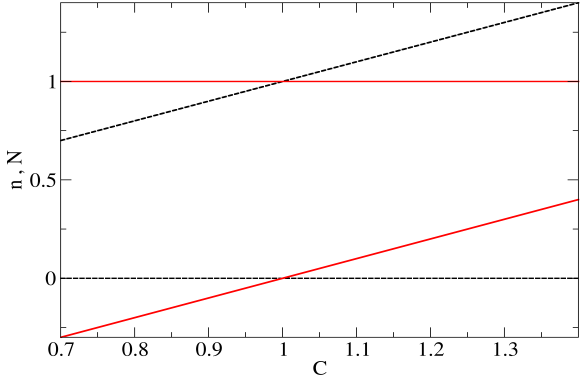

FIG. 13. Steady states for the two variables of the REs in the thermodynamic limit. The nonlasing solution is represented by the dashed lines (black online), the lasing solution by the solid ones (red online). The top pair of lines traces the dependence of the population $N_t$ as a function of pump rate $C$, the bottom one the photon number $n_t$.

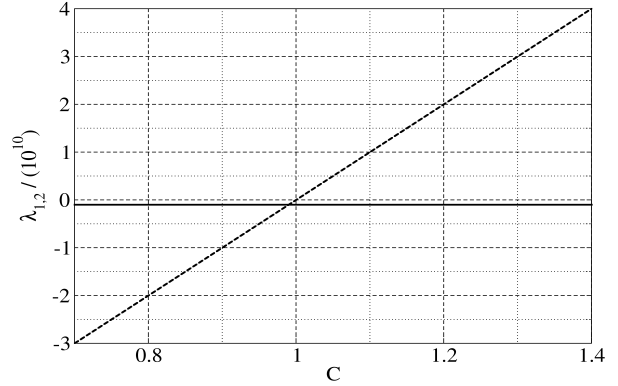

FIG. 14. Eigenvalues for the nonlasing solution. Notice that they are real over the whole interval shown (and even beyond). The nonlasing solution is destabilized at $C = 1$, when one of the eigenvalues becomes positive. It is therefore no longer a valid regime of operation for the laser.

In this limit, the presence of the spontaneous emission is neglected, since its contribution is several orders of magnitude smaller than that of its coherent counterpart even at threshold, thus permitting a particularly simple form of the model, written as[18]:

$$\dot{N}_t = -\gamma\left[(1+n_t)N_t - C\right] \,, \tag{5}$$

$$\dot{n}_t = -\Gamma_c(1 - N_t)n_t \,, \tag{6}$$

where the " ˙ " denotes the temporal derivative for the two variables ($n_t$ photon number and $N_t$ carrier, or population inversion, and the subscript $_t$ denotes the thermodynamic limit), $\Gamma_c$ stands for cavity losses, $\gamma$ for the intrinsic population relaxation rate and $C$ is the normalized pump rate.

This simple model, which can be directly derived from heuristic considerations (e.g., [70]), matches the conditions for which the equivalence between the laser threshold and a second-order phase transition hold [89–92]. Its properties are well-known and are briefly summarized here, for the benefit of later comparison. The stable laser operation depends on the time-independent solutions of the model – so-called steady-states – which tell what is the photon number to be expected (intracavity in eqs. (5,6) – the usable laser light is the fraction transmitted by the output coupler). The stability of each solution, for each pump rate value, tells whether the predicted photon number can be observed.

The steady state solutions – identified by the subscript $_s$ – can be easily obtained by solving eqs. (5,6) for $\dot{N}_t = 0$, $\dot{n}_t = 0$ and read:

$$\begin{pmatrix} {}^n n_{t,s} = 0 \\ {}^n N_{t,s} = C \end{pmatrix} \quad , \quad \begin{pmatrix} {}^l n_{t,s} = C-1 \\ {}^l N_{t,s} = 1 \end{pmatrix}, \tag{7}$$

where the lasing solution ${}^l n_{t,s}$ (thus also ${}^l N_{t,s}$) is physically meaningful only for $C \geq 1$ (below this pump value ${}^l n_{t,s} < 0$). The nonlasing solution ${}^n n_{t,s}$ (thus also ${}^n N_{t,s}$) is acceptable for any value of the pump rate $C$. Fig. 13 displays the well-known plot for these two stationary solutions: the dashed lines (black online) correspond to the nonlasing ($n_t$, $N_t$), the solid ones (red online) to the lasing ($n_l$, $N_l$) condition. Crossing of the solutions (for each variable) indicate the existence of a transcritical bifurcation [79], whose existence can be verified looking at the relaxation (or amplification) of an infinitesimal perturbation in the neighbourhood of each point ($n_t$, $N_t$).

This *linear stability analysis* [79] is governed by the eigenvalues

$$\begin{pmatrix} {}^n \lambda_{t,1} = -\gamma \\ {}^n \lambda_{t,2} = -\Gamma_c(1-C) \end{pmatrix} \quad , \tag{8}$$

$$\begin{pmatrix} {}^l \lambda_{t,\pm} = \frac{-\gamma C \pm \sqrt{\gamma^2 C^2 - 4\gamma\Gamma_c(C-1)}}{2} \end{pmatrix}, \tag{9}$$

for the nonlasing ${}^n \lambda_{t,1,2}$ and lasing ${}^l \lambda_{t,\pm}$ solutions, displayed in Figs. 14 and 15, respectively.

The nonlasing solution, Fig. 14, is stable and therefore experimentally observable for $C \leq 1$, since ${}^n \lambda_{t,1,2} \leq 0$ (both are fully real over the whole interval). Thus, below

---

[17] After decades of optimization, it has been possible to construct lasers, covering a large part of the near-InfraRed (and partly Medium IR), visible and near Ultra Violet, based on either semi-conducting materials or solid-state crystalline or amorphous structures. Most current micro- and nanolasers possess properties which belong to the dynamical class B. For better comparison, we thus concentrate on this category of devices even when considering their macroscopic equivalents.

[18] This form is adapted from [99], with pump rate $C$ normalized to its threshold value.

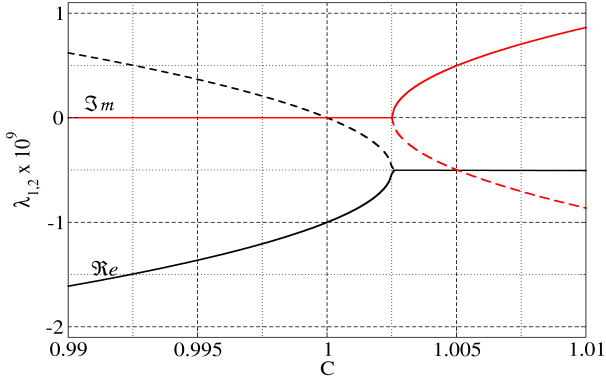

FIG. 15. Eigenvalues for the lasing solution. $\Re e$ identifies the real part (black online); and $\Im m$ the imaginary part (red online). Solid and dashed styles distinguish the real and imaginary parts of the eigenvalues when they differ (their superposition is recognized by the existence of a single solid line). The lasing solution is always unstable for $C < 1$ (one positive eigenvalue). Notice that for $1 \leq C \lesssim 1.0026$ the lasing solution is stable, but the eigenvalues are real, thus no coupled dynamics between photon and carriers is expected.

threshold there are no coherent photons inside the laser cavity[19].

The lasing solution, Fig. 15, instead is unstable for $C < 1$ (one eigenvalue has positive real part – black dashed line) but becomes stable for $C \geq 1$ ($^l\lambda_{r_{1,2}} \leq 0$). Thus, the intracavity photon number is $n_t > 0$ and coherent emission ensues. This is exactly what is described as laser threshold in terms of bifurcation theory. It is interesting to notice that both eigenvalues are strictly real for $C \lesssim 1.0026$, thus in the interval $1 \leq C \lesssim 1.0026$ any perturbation in the photon number is going to relax exponentially towards its steady state value $^l n_{t,s}$, without oscillations; in other words, no Relaxation Oscillations (ROs) occur in this pump range. The result has been known for a long time (e.g. [82]), but the pump rate range is so narrow as to be experimentally inaccessible. It is however interesting from a conceptual point of view, since ROs are considered a standard feature of class B lasers. For $C \gtrsim 1.0026$ (within any reasonable pump range [82]) the eigenvalues remain complex, pointing to a coupled photon number–population inversion dynamics. The real part of the eigenvalue (not visible in the expanded section of the pump rate axis in Fig. 15) steadily decreases with increasing pump, thus indicating a gradual reduction of the ROs as the operating point moves away from $C \approx 1.0026$.

Summarizing[20], the lasing threshold is well-identified in the context of this model and coincides with the change

_____________

[19] Recall that the spontaneous emission is neglected.
[20] Additional information, which can be used to compare to other features of the same model to which is added the average contribution of the spontaneous emission, is available in Section S-2.

in stability between the non-lasing and the lasing solution at $C = 1$.

## VI. RESE

We now turn to the analysis of the most basic model which has successfully been used to describe the physics of small-sized lasers. The RESE takes the form [6, 9], in a notation similar to that of Section V:

$$\dot{n} = -\Gamma_c n + \beta\gamma N(n+1), \tag{10}$$

$$\dot{N} = R - \beta\gamma Nn - \gamma N, \tag{11}$$

where $n$ and $N$ represent the photon and carrier number (or population inversion), respectively, $\Gamma_c$ and $\gamma$ are the relaxation rates for the intracavity photons and for the population inversion, respectively, and $R$ is the pump rate. The fraction of spontaneous emission coupled into the lasing mode is represented by the coefficient $\beta$ ($0 \leq \beta \leq 1$), where the lower limit matches the thermodynamic limit discussed in section V and the upper one corresponds to a device where only one electromagnetic cavity mode exists: the lasing mode (e.g., a simple nanocavity of half-wavelength size in the emitting direction, and smaller in the other ones, if we neglect the Purcell effect [50]).

Notice that while eqs. (10,11) are correct for any $\beta \neq 0$, the limit $\beta \to 0$ makes the coupling $\beta\gamma Nn$ vanish; this is the reason why the REs are written from the start in a slightly different form. The contribution of the average spontaneous emission scales correctly and in this case the limit $\beta \to 0$ holds. We will later show (Section VII B 1) that in spite of the formal differences, the RESE (eqs. (10,11)) correctly converges to the thermodynamic limit (eqs. (5,6)).

Notice that eqs. (10,11) – as the REs in Section V – mimic the behaviour of a 4-level system and are suitable for Quantum Well devices, as is the device used in the experiment (Section IV); a similar form can be written for Quantum Dots [52], which are formally closer to a 3-level system.

The steady state solutions for this model read

$$\langle n \rangle = \left\{ \left( \frac{C-1}{2} \right) + \sqrt{\left( \frac{C-1}{2} \right)^2 + \beta C} \right\} \beta^{-1}, \tag{12}$$

$$\langle N \rangle = \frac{\Gamma_c}{\beta\gamma} \frac{C}{1 + \beta\langle n \rangle}, \tag{13}$$

$$C = \frac{R}{R_{th}}, \quad R_{th} = \frac{\Gamma_c}{\beta}, \tag{14}$$

where $C = 1$ is formally considered to be the laser threshold [8] and is interpreted below (and in Section S-4 A). The normalization to the physical threshold pump rate $R_{th}$ enables comparison among different values of $\beta$; however, the limit $\beta \to 0$ can only be implemented with very small $\beta$ values but never replacing it with 0. At variance with the REs, only the solution for $\langle n \rangle$ with a positive

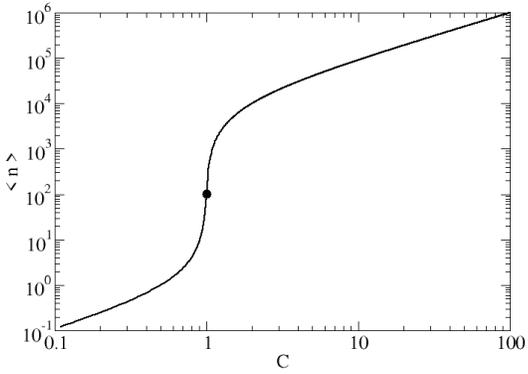

FIG. 16. Input-output laser response curve for $\beta = 10^{-4}$. The dot, marking $\langle n \rangle_{th}$, appears at mid-height between the lower and upper branch in the double-logarithmic representation, corresponds to the GLT (cf. Section S-4 A).The physical pump rate at threshold is $R_{th} = 10^{15} s^{-1}$ pump cycles (eq. (14)).

sign in front of the square root is considered here[21].

The "threshold" definition adopted here, $R_{th}$, corresponds to the balance between gain and losses, which attributes to the (average) photon number "at threshold" its sole spontaneous emission contribution, as can be easily seen by taking $C = 1$ in eqs. (12,13):

$$\langle n \rangle_{th} = \frac{1}{\sqrt{\beta}}, \quad \langle N \rangle_{th} = \frac{\Gamma_c}{\gamma_\| \beta} \frac{1}{1 + \sqrt{\beta}}.$$ (15)

It is useful to remark that $\langle n \rangle_{th}$ is nothing but the geometric average between the expected single photon per mode at the threshold point (QT) and the above threshold photon number $\beta^{-1}$ and matches the *practical recipe* of adopting the mid-point of the steep portion of the laser response as a "threshold" estimate.

### A. Predicted laser response

The first interesting comparison between REs and RESE concerns the laser response and is based on a simple analysis of the steady states. Two sets of solutions have emerged for the REs (Fig. 13), while the only physically meaningful one for the RESE is given by eqs. (12, 13) (Fig. 16, for $\beta = 10^{-4}$, as in the experiment[22]).

Aside from the double logarithmic scale, the difference between the two sets of solutions is rather striking. For the REs, the photon number *below threshold* [23] is strictly 0, thus the vertical logarithmic scale cannot be used, while above threshold a straight line emerges. From the RESE the more complex structure appears for $\langle n \rangle$, in close analogy with what measured experimentally (Fig. 2). The origin of this discrepancy is best understood by looking at an enlargement of the $C \approx 1$ region, where in the left panel of Fig. 17 we plot simultaneously the physical, $\langle n \rangle$ (solid lines, black online), and the non-physical, $\langle n_- \rangle$ (dashed lines, red online), solutions. The right panel shows the equivalent curves for the population inversion.

---

[21] The second solution is non-physical, as we see in Section VI A, and will be denoted $\langle n_- \rangle$.

[22] The steady state response curves for different $\beta$ values are plotted in Section S-3.

[23] The italics is used to signify that "threshold" is employed in a conceptual way, rather than following a strict definition.

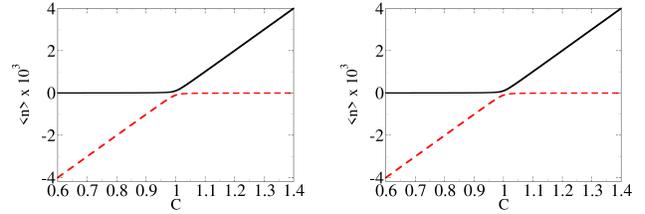

FIG. 17. Detail of the two steady state solutions of the RESE in a neighbourhood of threshold ($C = 1$). The left panel displays the stationary photon number value $\langle n \rangle$, the right one the corresponding value of the population inversion $\langle N \rangle$. The dashed solution (red online) corresponds to the non-physical one with the negative sign in front of the square root in eq. (12). Contrary to the transcritical bifurcation of the REs (Fig. 13), here the solutions remain distinct.

The avoided crossing in the two solutions, which visually resembles an avoided level crossing in quantum mechanics, stems from the average contribution of the spontaneous emission added to the coherent photon number $\langle n \rangle$ and gives rise to a so-called *imperfect bifurcation* [100–103]. This way, the **bifurcation disappears** and with it the possibility of defining a threshold outside the thermodynamic limit.

This is the origin of the difficulty in defining a laser threshold in the RESE, but also the source of interesting insights into the reasons why the observed dynamics (Section IV H) is more complex than a simple phase transition. On the one hand, the imperfect bifurcation follows directly from the choice of adding the average spontaneous emission to the laser field and represents the fundamental model's limitation (cf. Section IX for further discussion); on the other hand, however, it offers insight into some of the complexities of the transition between incoherent and coherent emission. It is important to remark, however, that the limitation is not a consequence of the simplicity of a rate-based approach, since a quantum statistical analysis [8] (which converges to the RESE) allowed for the extension of the rigorous LFT threshold definition (in the sense of the largest fluctuations) to lasers away from the thermodynamics limit. The predicted quantum-statistical threshold matches exactly the definition of GLT but, as already recognized [8], fails at smaller laser scales. This point will be later discussed in more detail in Section VIII.

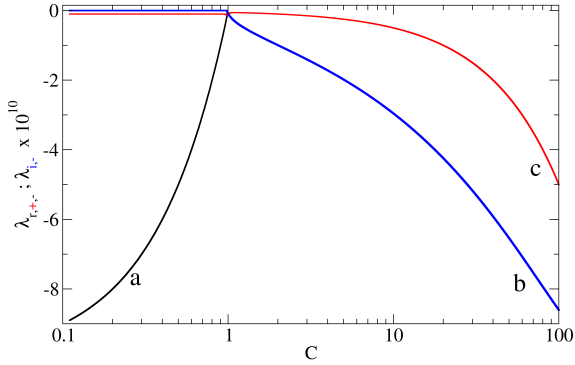

FIG. 18. Eigenvalues of the laser stability for $\beta = 10^{-4}$ plotted in semi-logarithmic scale for the physically relevant solution (solid lines in Fig. 17 and Fig. 16). The real parts are identified by the letters (a) for the most stable eigenvalue (black online) and (c) for the least stable one (red online); line (b) – blue online – represents the negative frequency component of the imaginary eigenvalue (the other one is identical, but with opposite sign). Notice that the eigenvalues' imaginary part is zero as long as the two real parts are distinct (in that region (a) and (c) overlap).

## VII. BRIEF TOPOLOGICAL ANALYSIS OF RESE

We first examine the case $\beta = 10^{-4}$, to best compare with the experiment, then enlarge the range to consider the whole interval of $\beta$ values spanning from the macro- to the nanoscale. Since only one solution has physical meaning, in the following we will entirely neglect the other one.

### A. Eigenvalues for a microlaser

As for the RE model (Section V), the stability of the solution ($\langle n \rangle$) and its dynamical properties are determined by the eigenvalues, plotted in Fig. 18. Curves (a) and (c) represent the real parts of the eigenvalues, while (b) shows the evolution of the negative component of the imaginary part (the other one is identical, with opposite sign).

For $C \lesssim 1$ the two eigenvalues are entirely real, distinct and always negative, confirming the statement that the $\langle n \rangle > 0$ is always stable; thus the solution is experimentally accessible, since it corresponds to a stable photon emission. One eigenvalue, however, is rather close to 0, signalling lower stability, while the second one is strongly negative (i.e., very stable). The two real eigenvalues merge into a complex conjugate pair (Fig. 19, left panel, $C \approx 0.97$) preceded by a rapid, partial gain in stability by the eigenvalue closer to 0. The imaginary part (positive branch) is displayed in the right panel of the same figure and sharply grows, indicating the appearance of coupled oscillations between the photon number and the population inversion in response to a perturbation.

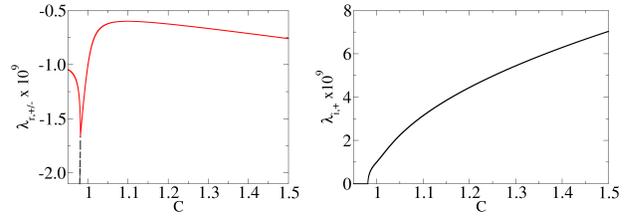

FIG. 19. Detail of the eigenvalues of the laser stability for $\beta = 10^{-4}$ plotted in semi-logarithmic scale for the physically relevant solution (solid lines in Fig. 17 and Fig. 16). The left panel represents the real parts, where the dashed line (black online) represents the most stable and the solid (red online) the least stable one. The right panel illustrates the evolution of the imaginary part (positive component, here) in the same interval.

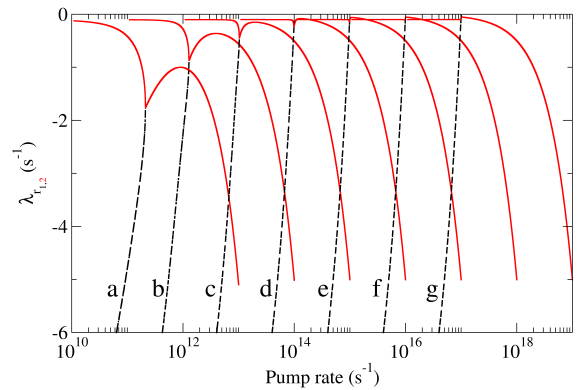

FIG. 20. Real part of the eigenvalues of the laser stability; the dashed lines (black online) correspond to the first, the solid ones (red online) to the second, less stable eigenvalue: at the cusp the eigenvalues form a complex-conjugate pair as in Fig. 19. $\beta = 1$ (a), $10^{-1}$ (b), $10^{-2}$ (c), $10^{-3}$ (d), $10^{-4}$ (e), $10^{-5}$ (f), $10^{-6}$ (g).

### B. Eigenvalues: general case

Fig. 20 shows the real part of the two eigenvalues going from the extreme nanolaser ($\beta = 1$) to a macroscopic laser ($\beta = 10^{-6}$). Two pieces of information emerge from the picture: 1. the overall behaviour remains the same, irrespective of laser scale, and 2. nanolasers are dynamically more stable (more negative real eigenvalues) than their micro- and macroscopic counterparts. The action of a nonnegligible amount of noise, stronger at the nanoscale, could have instead suggested an intuitive picture of reduced stability.

Fig. 20 belies this idea highlighting the difference between noise and dynamical stability: the latter determines the speed at which the system relaxes back to its operation set point, irrespective of the number and amplitude of perturbations to which it is subjected, which are controlled by the former. Thus, there is no incompatibility between larger noise and better stability and nanolasers turn out to be more stable. This property is also a potential advantage in telecommunications, promising reduced error rates for the same amount of optical power [33].

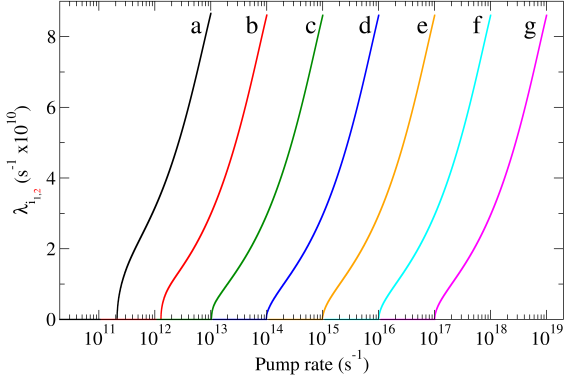

FIG. 21. Imaginary part of the eigenvalue (positive, the negative one is the mirror image with respect to the horizontal axis) of the laser stability. $\beta = 1$ (a), $10^{-1}$ (b), $10^{-2}$ (c), $10^{-3}$ (d), $10^{-4}$ (e), $10^{-5}$ (f), $10^{-6}$ (g).

The results of the *lsa* also explain the very early remark [6] that nanolasers do not have ROs. In the absence of a stability analysis, this fact could be attributed to an "unknown" physical property which decouples the dynamics of photons and carriers. Instead, it simply amounts to a strong damping of the oscillation amplitude in a single cycle[24], as can be seen by comparing the oscillation frequency coming from $\lambda_i$ and the damping rate ($\lambda_r$).

The positive imaginary part of the complex conjugate eigenvalues is shown in Fig. 21. The behaviour is substantially the same for all kinds of lasers, except for the $\beta = 1$ case where the oscillation frequency grows more sharply in the initial phases. More details on the evolution of both the real and imaginary parts of the eigenvalues are given in Section S-4 F.

### 1. Convergence towards the thermodynamic limit

One important point to address is the convergence of the RESE to the REs (thermodynamic limit). Fig. 22 reproposes the information of Fig. 20 with the normalized axis on the horizontal scale, for better comparison. The least stable eigenvalue, traced in Fig. 22 for a broad range of $\beta$ values, clearly shows the evolution from the nanolaser's stronger stability to the macroscopic one's reduced robustness to noise. We also remark that for each $\beta$

---

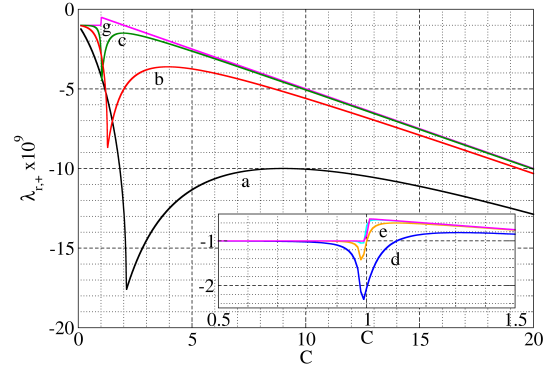

FIG. 22. Real part of the least negative (i.e., least stable) eigenvalue plotted as a function of $C$ for different values of $\beta = 1$ ((a), black online), $10^{-1}$ ((b), red online), $10^{-2}$ ((c), green online), $10^{-6}$ ((g), purple online). In the inset, from bottom to top, we find $\beta = 10^{-3}$ ((d), blue online), $10^{-4}$ ((e), orange online), $10^{-6}$ (purple online). The curve matching $10^{-5}$ (cyan online) is nearly entirely covered by the purple line.

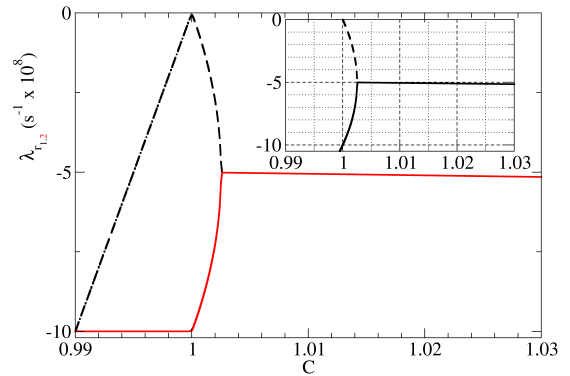

FIG. 23. Detail of the real parts of the two eigenvalues for $\beta = 10^{-9}$ in a small neighbourhood of the "threshold". The inset shows, for comparison, a detail from the corresponding eigenvalues taken from the REs (thermodynamic limit, cf. Fig. 15 for a larger scale).

there is a value of pump for which the complex-conjugate pair of eigenvalues (to the right of the cusp) has a maximum real part (lowest stability) and that the corresponding pump value moves towards $C = 1$ as $\beta$ decreases. The eigenvalue's approach towards 0, from below, as $\beta$ decreases indicates a progressive convergence towards the thermodynamic limit, which is best illustrated by the pair of eigenvalues (real parts) plotted in Fig. 23 for $\beta = 10^{-9}$. Here, the closeness of $\lambda_{r1}$ to 0 is quite obvious, as is its evolution with $C$. The inset shows a detail (same interval as main figure) of the real parts of the eigenvalues for the REs (Fig. 15). The match for $C \geq 1$ is excellent even at a quantitative level, showing that the RESE do converge correctly to the REs above $C = 1$. However, for $C < 1$ the RESE possess two negative eigenvalues, while the REs have one positive one (cf. Fig. 15). Notice that the two real eigenvalues merge into a complex-conjugate pair at $C \approx 1.0026$ in both models. We therefore conclude that the RESE is a reliable way of seamlessly connecting the thermodynamic limit to all finite-size lasers.

---

[24] At a pump rate $\sim 10^{12} s^{-1}$ excitations per second, matching the smallest damping, the oscillation amplitude is reduced to approximately $\frac{1}{3}$ in a single period (cf. Figs. 20,21). For all other pump values damping is even stronger!

## VIII. INTERPRETATION OF THE TOPOLOGY

The previous observations contain interesting implications for the evolution of threshold as a function of laser scale. Out of the eight different definitions of threshold that we have chosen to examine, six are directly interpretable in terms of these results[25].

We will choose to relate the findings to the GLT, whose nature is related to the photon number, thus easily identifiable and already considered (e.g., [8]). As already seen in the experiment, its position coincides with the geometric mean of the average photon number values taken on the lower and on the upper emission branches (Section S-4 A) for all values of $\beta$ (except, of course $\beta = 1$, for which the position cannot be determined: the *thresholdless* laser).

For the other threshold definitions, the following factors emerging from the previous considerations contribute to their traits:

1. The *lsa* shows that the laser stability depends on its $\beta$ value (loosely connected the its inverse cavity volume): the larger $\beta$, the more stable the device. Hence, the most unstable devices are those that match the thermodynamic limit for which an actual threshold, as phase transition, can be defined in a statistical physics sense.

   Three different aspects of the stability contribute to the different (accessory) threshold definitions:

   (i.) the absolute value of the relaxation rate ($\Re e \{\lambda\}$) over the whole pump interval and, in particular, its maximum value $\lambda_{r,max} = max \{\Re e \{\lambda\}\}$ (recalling that $\Re e \{\lambda\} < 0, \forall C$);

   (ii.) the "least stable" pump value: $C_{ls} = C(\lambda_{r,max})$;

   (iii.) the extension in pump interval over which the eigenvalues remains closest to 0 for $C \geq 1$.

   Notice that $1 \leq C_{ls} \lesssim 10$ with a strong nonlinear dependence as $\beta$ spans the $(0, 1]$ interval (Fig. 22).

2. Reduced stability lessens the damping of fluctuations away from the equilibrium point (eqs.(12, 13)), thus permitting larger deviations.

3. The actual largest observable fluctuation is a result of the combination of the three factors (i.)...(iii.).

4. Absolute fluctuations grow with pump value (both in the experiment, Section IV F, and in the RESE, Section S-4 G), with an asymptotic growth $\propto n^{3/2}$ (eq. (S-9)).

5. The "jump" in photon number when passing from the lower to the upper branch is $O(\beta^{-1})$.

The coincidence of (i.) [largest $\lambda_{r,max}(\beta)$ (for $\beta \to 0$)] with (ii.) [$C_{ls} = 1$ (or nearly so)] and (iii.) [a very peaked functional dependence in $C \approx 1$ for $\lambda_r(C)$] produces the traditional laser threshold which holds for macroscopic lasers, with the traditional features of a phase transition. Furthermore, it is the coincidence of these three factors at one pump value ($C \approx 1$) which enables the equivalence among most of the definitions we have considered so far: the statistically defined thresholds LFT, PST[26], CT and FT all converge towards the GLT (which always takes place at $C = 1$). In the absence of a simultaneous occurrence of these factors, the different functional dependence on fluctuations and average photon number of the statistically defined thresholds would separate their predictions.

At the opposite extreme ($\beta \to 1$) $|\lambda_{r,max}(\beta \approx 1)| \gg \lambda_{r,max}(\beta \approx 0)$ (strong damping!) couples with a broad pump range ($\Delta C \approx 4$), centred around $C_{ls} \approx 10$, where stability is weakest. This results in smaller fluctuations which take place at much larger pump values (than in the macroscopic regime) and extend over a large pump interval. The differences in functional dependence of the threshold indicators for this $\beta$ value do not produce a concomitance, but rather a competition of factors which determine the threshold, leading to diverging predictions from one definition to the other.

These observations will be applied to one specific case, the FT, in Section VIII A, while the other thresholds will be examined in Sections S-4 A through S-4 G. The notable exception to the previous discussion is the ROT, with its exclusive dependence on dynamical properties (similarly to the BT) since it is identified by the transformation of the two real into a complex-conjugate pair of eigenvalues. While in the limit $\beta \to 0$ we find an approximate coincidence of the ROT with all other definitions (up to the small deviation shown in Figs. 15 and 23 – the conjugate pair occurs at $C \approx 1.0026$), the picture becomes much more complex as $\beta$ progresses from 0 to 1. In particular, as shown in Figs. 20 and 22, and discussed in more detail in Section S-4 F, there is an intermediate region at the border between the macroscopic and the microscale where an intricate, non-monotonic dependence of the ROT on $C$ is observed. The complexity of these details does not warrant further analysis in this paper.

---

[25] We exclude the BT – since no bifurcation exists in the RESE (intrinsic shortcoming) –, and the QT whose discussion we have decided to forgo because of its more complex nature which requires a dedicated investigation.

[26] It is useful to remark that the following expression

$$g^{(2)}(0) - 1 = \frac{F - 1}{\langle n \rangle}$$

directly links PST and CT to the FT through the fluctuations (governing the LFT).

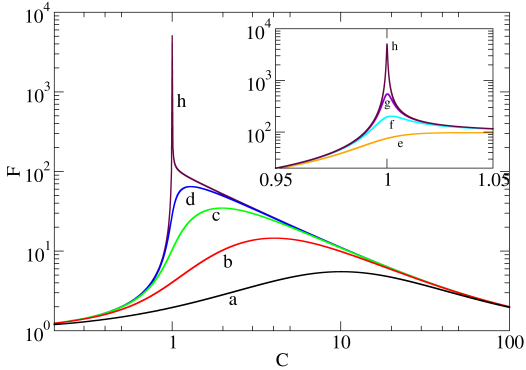

FIG. 24. Fano function, computed from RESE, for $\beta = 1$ (a), $10^{-1}$ (b), $10^{-2}$ (c), $10^{-3}$ (d), $10^{-4}$ (e), $10^{-5}$ (f), $10^{-6}$ (g), $10^{-8}$ (h). The inset shows a detail of the cusp highlighting the change in shape of the Fano function between macro- and microlasers.

### A. RESE Fano Threshold

Plotting the Fano function[27] for the ensemble of $\beta$ values, we remark strong differences in their general features (Fig. 24): the expected sharp peak [8], which gives a clear threshold identification, is very well identifiable in a large macrolaser ($\beta = 10^{-8}$, curve (h)); its presence is, however, recognizable for all macroscopic lasers ($\beta \leq 10^{-5}$) as displayed in the inset of Fig. 24. A relatively sharp maximum, devoid of cusp-like shape, appears for microlasers (curves (d) and (e) in Fig. 24), which however transforms into a shifted, broad and ill-defined feature in nanolasers. The position of the peak shifts towards ever increasing values of $C$ as the laser size shrinks (for $\beta = 1$ the maximum occurs at $C \approx 10$).

Fig. 24 clearly summarizes the considerations 1-5: the strong damping which characterizes nanolasers reduces the amplitude of the fluctuation, thus the amplitude of the Fano factor, with a maximum which appears at large pump values. Hence, this indicator quantifies the photon variance relative to the average photon number[28], but cannot possess any meaning in terms of laser threshold, not only due to lack of information on the growth of coherence, but also because the largest fluctuations are the result of the combined factors enumerated at points 1-5. Thus, macroscopic lasers aside[29], the FT cannot be considered to be a *threshold* in any sense, but just a measure of the fluctuations. The evolution from the nano- to the microscale is gradual and consists of a shift towards lower $C$ values and a gradual sharpening of the maximum.

### B. Additional remarks

Commenting on some additional aspects which emerge from the RESE analysis, we can establish some additional parallels and differences with the dynamics of the threshold transition emerging from the experiment. The first aspect, which differentiates observations and RESE modelling, concerns the report on the emergence of photon bursts (Section IV H), which cannot be found from the analytical predictions. As already mentioned, bursts are compatible with the topological structure of the RESE phase space [54, 56], but are not part of the predictions of analytical models (except for the *ad hoc* modification introduced in [61], whose ability to produce bursts has been reconduced to the way it affects the phase space structure [54]). Yet, comparing the largest measured fluctuations (Fig. 6, and Fig. S-24 for the experimental average fluctuations) to the RESE functional dependence (Fig. S-22) one may be induced into thinking that the qualitative similarity of the curves should suggest a match between experiment and model. This cannot be the case, since the experimentally observed bursts belong to an emission regime that can be assimilated to ASE [54], with superthermal photon statistics [59], while the RESE prediction is accompanied by subthermal statistics. A comparison between the two regimes is therefore not possible.

The displacement of the maximum in the real part of the complex eigenvalue away from $C = 1$ (Fig. 18 for the microlaser and Fig. 22 in general)) has already explained the observed lack of coincidence in threshold predictions using different definitions. Now it also explains the experimental observation of the stronger ROs at a point beyond most threshold definitions (Fig. 12), which are observed at $i = 1.60mA \approx 1.18i_{GLT}$ (Fig. 10c). Comparison to Fig. 18 (left panel) shows that the lowest stability occurs at $C \approx 1.1$; it is around this pump value that one should expect the smallest damping for the ROs, thus the strongest spectral rf component. Notice that the numerical value is close enough to the experimentally determined pump value[30] (in units of GLT).

Finally, we remark on the compatibility in the pump values at which convergence of $g^{(2)}(0)$ towards Poissonian statistics is reached in the experiment (Fig. 3) and in RESE (Fig. S-16). The lack of error bars in the numerical predictions, contrary to the experiment [59], does not allow for the precise determination of the pump value for which Poisson statistics is attained, but choosing a reasonable range (e.g., $g^{(2)}(0) = 1.01$, or 1% above the theoretical value) provides an acceptable estimate. Additional comments on the behaviour of $g^{(2)}(0)$ are given in Section S-4 B).

---

[27] For this, we use the expression for the photon number variance given in eq. (S-9), which can also be derived from [52, 104]. We are grateful to J. Mørk for the expression of eq. (S-9).

[28] It is useful to remark that most of the curves in this paper are plotted as a function of normalized threshold $C$ (eq. (14). The photon number differs greatly from one value of $\beta$ to another (cf. Fig S-12), in spite of the same $C$ value.

[29] For $\beta = 10^{-5}$ the peak of the Fano factor is located at $C \approx 1.002$, close enough to be considered nearly coincident with the GLT.

[30] Recall that the $\beta$ value used in RESE for the microlaser is only close, but not identical to the experimentally inferred range.

## IX. ASSESSMENT AND CRITICISM OF RESE

The RESE model has the indubitable merit of providing a wealth of information matching experimental observations and conceptual considerations. It also possesses the additional advantage of being derivable from intuitive considerations, based on fundamental properties of radiation-matter interaction [105], as well as from models soundly grounded in electromagnetism as well as on quantum mechanics [8, 70, 98, 106]. A further benefit comes from its intrinsic simplicity, since a description based on two dynamical variables (population inversion and photon number) offers the possibility for direct insight into many physical properties. The discussion has also shown that other unexpected features find a match with observable quantities or conceptual properties, thus increasing the RESE usefulness. At the same time, since all modelling choices are subject to approximations and thus possess shortcomings, it is important to highlight some of the problems which are intrinsically rooted in the form of the model itself.

In contrast with the REs (valid in the thermodynamic limit, thus devoid of spontaneous emission, eqs. (6,5)), the RESE introduce the average of the background noisy field into the lasing mode. While, on the one hand, this choice palliates the extreme assumption of the thermodynamic limit – unsustainable for small lasers – it does so by adding the average contribution of the **spontaneous emission** (thus incoherent field) to the **coherent photon number** $n$. Thus, it implicitly *imposes an inexistent phase agreement between the spontaneously emitted photons* (random variable) *and the coherent ones resulting from stimulated emission* (deterministic variable). This paradoxical standpoint, simultaneously strength and weakness of its straightforward description, is the source of the problems illustrated below.

It has already been shown that the structure of the bifurcation changes entirely from the REs to the RESE. Regardless of its size, $\beta$ breaks the transcritical bifurcation transforming the two crossing branches into two separate ones (imperfect bifurcation [100, 103] characteristic of the RESE. In addition, one branch (physically meaningful) is always stable, the other (non-physical) is always unstable (Fig. 17): the bifurcation has thus disappeared leaving behind a (physical) solution stable for any value of pump. With the disappearance of the bifurcation, the BT has also disappears and with it the possibility of defining the threshold concept in a sound way.

Looking at the problem more physically, we see that the addition of (pseudo-)coherent photons to the coherent field renders the latter always positive for all pump values: the laser appears to emit at all times a certain amount of "coherent" light. Although one can *interpret* the result by considering that "below threshold" the photons are incoherent, there are two clear shortcomings: first, the mathematical tools that can be employed to find threshold can not make the distinction and therefore break down, and, second, if the visual identification fails (as in the case of a large $\beta$ laser) it is impossible to identify the pump range in which the emission would be incoherent. Worse, in the presence of a pump region in which both kinds of emission coexist (as in small lasers [54–56]), then the picture becomes entirely blurred, leading to the confusion that has characterized the concept of laser threshold in small devices for the past decades.

One additional problem is represented by the use of continuous variables, necessary in a differential representation, but non-physical in nature. This is not a peculiarity of the RESE, since all differential descriptions suffer from the same shortcoming. However, the problem becomes particularly serious when relatively small photon and carrier numbers are considered [107], as in micro-and – especially – nanolasers. The argument used in the thermodynamic limit that the addition or removal of one photon is negligibly small compared to the macroscopic (coherent) photon number fails when the photon number is small. A discrete modelling offers in this respect highly superior performances [107], particularly because it can intrinsically include granularity and noise without further assumptions [60, 61, 87, 88].

### A. How can the laser threshold be properly modelled

The actual physical threshold, which corresponds to the growth of a coherent field out of nothing, requires the presence of a phase, thus cannot be described with a photon number; a (preferably quantized) field is needed. For such a description, the coherent field (possessing thus a well-defined phase) must be defined also "below threshold" although it will be equal to zero until the bifurcation occurs. These are the conceptual lines along which a new description of lasing, valid at all scales, has been recently developed [65]. However, this requires a larger phase space (7 variables) and is therefore less practical. The usefulness of the RESE holds, therefore, as long as one understands correctly its features and limitations, and thus interprets its predictions in the correct way. The discussion on thresholds presented in this paper aims at fulfilling this role.

It is important to understand that a bifurcation is indispensable for a proper description of laser threshold and that ***its absence is not a feature of small-sized lasers***, for which the bifurcation still exists, but only a <u>model limitation</u>[31]. Recognition of this crucial property helps putting in perspective different modelling choices and correctly interpreting the ensuing predictions. In other words, while the RESE offer an excellent insight into many laser properties at all scales, nonetheless, they also provide conceptually erroneous results when trying to extract the threshold point. The shortcoming does not originate from a **peculiarity of small lasers**, but by the nonexistence of the essential feature which physically defines threshold (bifurcation) and by the erroneous supposition that the equivalence of secondary definitions, such as those discussed in this paper, holds at all laser scales. Interestingly, many of the physical properties experimentally observed are reproduced by the RESE[32], thus confirming its extreme usefulness and its remarkable predictive abilities.

---

[31] It has been recently suggested that bifurcations would not appear in a properly elaborated model [108]. Independently of the mistakes made in the proof of the statement in [108], a laser model which properly describes the growth of a coherent field <u>requires</u> the existence of a bifurcation.

### B. Why is the RESE so successful

The reason for the success of the RESE resides in having captured an essential point underlying the physics of lasers at all scales: the contribution of the spontaneous emission, inversely proportional to a generalized "cavity volume"[33], adds a stabilizing contribution to the interaction between photons and carriers. When $\beta$ is very small, this term is nearly negligible and represents only a small correction to the description of the REs; indeed, the RESE correctly converge towards the thermodynamic limit as $\beta \to 0$ (Section VII B 1). In this regime, the sharp growth of the photon number originates from the bilinear term ($\beta \gamma N n$ in eqs. (10, 11)) which describes the joint action of carriers and photons, thus stimulated emission.

When, instead, $\beta$ approaches 1, the term representing the spontaneous contribution $\beta \gamma N$ supplies a source for the growth of the photon number linearly proportional to the carrier availability, thus reducing the influence of the bilinear term, whose signature is an abrupt growth. As a result, when progressing from the macroscopic to the nanoscopic scale, the sudden growth of the photon number is progressively reduced to mirror the weight of the relatively weaker effect of the nonlinear gain process.

Expressed in more heuristic and physical terms, we can picture the two extreme cases as follows. In the macroscopic case, a large amount of spontaneous emission is lost into the electromagnetic cavity modes which do not contribute to lasing. Since the dominance of the stimulated process requires a minimal amount of photons (one, averaged over time, according to the QT concept) and given that below threshold the photons can be supposed to be homogeneously distributed over all modes, threshold requires a large pump rate (cf., e.g., Figs. S-11 and 20). However, as soon as the stimulated emission avalanche starts, the photons which were otherwise emitted into the non-lasing modes are suddenly "redirected" towards lasing, thus resulting into a large and very sharp jump. This way, threshold is clearly identified.

For small cavities, the number of electromagnetic field modes is also quite small ($\approx \beta^{-1}$). As a result, a smaller amount of energy is lost into spontaneous emission and threshold occurs sooner (cf., again Figs. S-11 and 20), but, at the same time, the number of photons which can be "redirected" into lasing is now strongly reduced. Combined with the smoothing action of the linear (spontaneous) process compared to the nonlinear (stimulated) one, one obtains the well-known results.

Favoring a heuristic imaging, we can state that the large electromagnetic reservoir proper of macroscopic lasers is responsible for the sudden switch from "isotropic" to "directed" (or lasing) emission. Small devices have comparably smaller electromagnetic reservoirs and cannot quickly provide a large reserve of photons to produce sudden lasing.

These images are useful and help understanding the reasons why the transition is more or less sharp, and also why the RESE is such a successful model. However, they do not tell us anything about the process which renders the emergence of a coherent field possible, out of nothing. The latter process, which corresponds to the essence of threshold, can only be described with the help of coherent fields and bifurcations.

### X. CONCLUSIONS

The RESE is an excellent model for lasers at all scales and is capable of predicting a wealth of information that matches the experimentally observed microlaser features. The same is expected to hold for nanolasers, even though at the present time fewer experimental results are available, for comparison. Nonetheless, the very good match between general dynamical observations in both micro- and nanolasers and the predictions based on RESE-like models lend further credibility to the soundness of this simple modeling approach [78, 109–120].

The success of the RESE rests on its sound description of the phase space in which the two physical variables interact. This also allows for a good characterization of the fluctuations and, as we have shown in this paper, contributes to explaining why the secondary threshold definitions disagree with one another (and with the BT). Nonetheless, the model fails to predict the bifurcation, due to the modeling choice of the spontaneous emission with its imperfect bifurcation structure [100–103].

Good threshold definitions require the identification of the BT and for this both coherent and incoherent fields have to be defined on both sides of the transition (cf., e.g., [65]). However, these models are much more complex and should be used only whenever necessary – at the present time they do not replace the RESE for the simpler predictions. It must be kept in mind that other modeling approaches have been successfully used to predict, for instance, particular features of the photon statistical properties of nanolasers or other dynamical aspects (cf. among others, e.g., [42, 51, 121–126]), but, again they cannot compete with the RESE for simplicity.

The results of this investigation can be summarized in the following points:

- Only the Bifurcation Threshold (BT) is scale-independent and holds for any kind of laser, regard-

---

[32] The photon bursts probably disappear due to the Langevin noise hypothesis, which amounts to applying small perturbations very often. However, this point requires further investigation.

[33] The detailed physics is more complex since it involves the Purcell effect, etc. For a recent discussion see [50].

less of its size. As such, it has to be considered as the **primary definition** of laser threshold.

- The physical transition from incoherent to coherent emission **requires** the existence of a BT; the absence of a BT only indicates a shortcoming of a model, irrespective of its other merits (e.g., RESE), and **not** a feature of particular lasers.

- Laser threshold is **not necessarily equivalent** to a sudden jump in the emission in the lasing mode, nor does it correspond to a rapid transformation of the photon statistics from thermal to Poissonian. The latter properties hold only in the thermodynamic limit (thus, to a good extent for macroscopic lasers). The absence of these features does not indicate a *faulty threshold* nor a problematic case; the difficulties in interpretation only come from the application of concepts used in large lasers to small devices. Abandoning this biased point of view permits a fully consistent threshold definition and a conceptually sound description of the process which describes the emergence of a coherent field.

- The accessory threshold definitions[34] (cf. Table I) are useful only in the thermodynamic limit (i.e., sufficiently large lasers) where they coincide **both** with one another, **and** with the BT. Instead, their predictions disagree when applied to micro- or nanolasers due to their different dependence on photon number and on the fluctuations' amplitude. Since the latter is affected by the contraction rate of perturbations (cf. the linear stability analysis of RESE, Section VII) which changes both in amplitude and in functional dependence with pump, it is obvious that the accessory definitions cannot be used in place of the primary one (BT).

- The Relaxation Oscillation Threshold (ROT) is based on a topological feature strictly related to the specific properties of any laser, irrespective of its size. Although, in general, it does not match the pump value at which the BT takes place, nonetheless, it possesses very interesting features which qualify it as a valid surrogate for the BT (unless more complex measurements are undertaken, cf. Section S-1 D):

  - Guarantee that only the coherent field is detected (the incoherent one cannot participate in the quadrature oscillations with the carriers, typical of the interaction between photon number and population inversion, cf. Fig. 1 in [127]). Thus, it is possible to identify the presence of coherent photons (even if in partial number compared to the total output) as long as coherence is established on long time scales (not on isolated pulses [54, 56]). This point is of no importance at the macroscopic scale, where photon bursts are expected to exist in a negligibly small pump range, but crucial for smaller devices.

  - Experimental accessibility. As already shown in experiments and confirmed by numerical predictions, a sensitive way of detecting the presence of the ROT in cw-pumped lasers is to modulate the injected current (or, possibly, the optical excitation) at a frequency close to resonance [86]. For pulsed lasers, a suitable reconstruction of the phase space also leads to a good discrimination of the coherent oscillations [71], thus of the ROT.

  - Use of rf spectra. While the previous techniques will generally lead to increased sensitivity in the ROT detection, rf spectra also contain the information and offer an even simpler analysis tool. The only shortcoming may be a lower sensitivity and/or increased difficulties in identifying the correct spectral feature.

  - Potential for extensions: short perturbations in cw or long-pulse system (similar to what known in macroscopic lasers – cf. response to switch in Fig. 1 of [128]) could be considered as alternatives to [86] for ROT detection.

- Except for its crucial role in macroscopic lasers, the PST is not to be considered a *threshold indicator*, but, rather, a quantitative indicator of the degree of coherence of the emitted light: its true role [26]. In smaller lasers, $g^{(2)}(0)$ can be used to determine the sufficient degree of coherence necessary for particular applications[35].

To summarize the physics at the transition as a function of laser size, we can reiterate the heuristic picture already used. As the ensemble of electromagnetic cavity modes act as an *energy reservoir* : a smaller reservoir (larger $\beta$) implies a reduced energy availability and reduced susceptibility to external perturbations. This picture can also be used to understand the smoothing out of the transition between incoherent and coherent emission.

In the thermodynamic limit, once the transition pump value is reached, the laser has an "infinite" amount of energy stored in the non-lasing electromagnetic cavity modes at its disposal[36]. At that point, any fluctuation is capable of driving the system away from the homogeneous distribution (QT: one photon per electromagnetic cavity mode, in average) towards the non-equilibrium one represented by lasing (all photons into the lasing mode): coherent emission ensues automatically thanks to the stimulated emission process [105].

---

[34] With the exception of the ROT, discussed in the following point.

[35] The already mentioned example of the threshold current declared by the manufacturer of the VCSEL-980 used in the experiment is an excellent implementation of this concept: the micro-VCSEL is considered to emit sufficiently coherent light typically around $i = 2.2mA$, with an upper limit at $i = 3.0mA$. This would match a range in autocorrelation values $g^{(2)}(0)_{[2.2mA]} \approx 1.4$ to $g^{(2)}(0)_{[3.0mA]} < 1.1$, as shown in Fig. 6 of [59].

[36] This is the reason why it is not possible to plot to scale the transition in Fig. 1: in this limit the upper branch has an infinite number of photons! Although it is a physical paradox, since it requires and infinite amount of energy, it correctly matches the statistical physics requirements for a phase transition [89–92].

At the opposite end of the scale, the *thresholdless laser* possesses only one electromagnetic cavity mode: the transition between thermal and lasing emission *unfolds* into the complete sequence of dynamical features. At intermediate $\beta$ values, the threshold physics gradually evolves between these two extrema, compressing all the steps (for larger lasers) into such a small pump interval as to render them experimentally inaccessible.

Irrespective of the width of the pump interval in which coherence emerges, the growing contribution of the coherent field component increases its relative importance in the total field to gradually reach dominance. This concept was already taught in laser courses[37] as an average, statistical concept. The new understanding which surfaces from the field of micro- and nanolasers completes the picture by adding the temporal dimension to the development of coherence [54].

As a final remark, the discussion on different indicators for the *phase transition*, presented here in the context of small lasers, actually applies to a much broader range of physical systems. In fact, the considerations on the different indicators' noise dependence – affecting the validity of their predictions – are entirely general and transfer to other structures and devices as their size shrinks. Thus, it is generally to be expected that difficulties will arise in the determination of the phase transition when using secondary indicators, while bifurcation theory will provide reliable information, irrespective of system size. In light of apparent misinterpretations of the role of bifurcations [108], it is important to reiterate the fact that a bifurcation signals the growth of a variable (the field amplitude in the case of a laser) out of zero, without excluding the simultaneous presence of a background – the incoherent intensity due to spontaneous emission. As such, the presence of the bifurcation **does not** amount to a sudden jump in the total photon number, which **can still linearly grow with pump** as a superposition of spontaneous and stimulated photons in variable proportions [65]. Thus, only a specific detection (ROT or the techniques described in Section S-1 D) will be able to experimentally identify the presence of the bifurcation.

**Note added in proof:** During the review process we have become aware of published work where fitting the experimental data with an analytical function [131, 132] avoids the noise-induced problems (Section IV G, Fig. 8) affecting the indicators proposed in [19].

### ACKNOWLEDGMENTS

This paper is the result of several years of work and, as a consequence, has benefitted from discussions with numerous colleagues. Thus, it is impossible to explicitly mention occasional discussions, even when important. Among colleagues with whom we have had the chance of exchanging ideas, we are particularly grateful to T. Ackermann, D. Aktas, O. Alibart, G. Almuneau, A. Beveratos, M.A. Carroll, L. Chusseau, G. D'Alessandro, L. Gil, J. Mørk, C.-Z. Ning, G.-L. Oppo, F. Papoff, I. Robert-Philip, É. Picholle, A. Politi, S. Reitzenstein, S. Tanzilli, and A. Yacomotti. Funding and support has been provided by the Région PACA and by "Investments for the Future" programme under the Université Côte d'Azur UCA-JEDI project managed by the ANR (ANR-15-IDEX-01). T.W. thanks the National Natural Science Foundation of China (Grant No. 61804036), Zhejiang Province Commonweal Project (Grant No. LGJ20A040001).

---

---

[37] F.T. Arecchi used to teach in his Laser Physics Lectures that the transition between the non-lasing and the lasing state occurs through a statistical superposition of spontaneous and stimulated photons, with coefficients which evolved in time during threshold crossing. This point of view – still valid, but enlarged by the new knowledge in the emission around threshold – stemmed from the transient experiments and the fact that in macroscopic lasers the unfolding of the threshold nowadays observable in smaller devices could only be seen as a temporal transition between the incoherent, subthreshold, and coherent, above threshold emissions [129, 130].

# Supplementary Material
## "Phase Transitions" in small systems: why standard threshold definitions fail for nanolasers

G.L. Lippi

*Université Côte d'Azur, CNRS, Institut de Physique de Nice*

T. Wang*

*School of Electronics and Information, Hangzhou Dianzi University, Hangzhou 310018, China*

G.P. Puccioni

*Istituto dei Sistemi Complessi, CNR, Via Madonna del Piano 10, I-50019 Sesto Fiorentino, Italy*

**CONTENTS**

## S-1. COMPLEMENTS ON THE EXPERIMENT

This section contains details about empirical techniques for threshold extrapolation in experiments (Section S-1 A), on the usefulness of threshold indicators proposed in [1] when applied to experimental data (Section S-1 B), and on the experimentally obtained Relative Intensity Noise function (Section S-1 C) which has not been shown in the main paper. Finally, we comment on two possible techniques for the direct identification of the Bifurcation Threshold (BT) in Section S-1 D.

---

* wangtao@hdu.edu.cn

## A. Empirical threshold extrapolation

A common technique used for obtaining a threshold estimate in microlasers is based on the extrapolation of the linear growth of the laser output as a function of pump. Fig. S-1 illustrates the procedure on the basis of the computed intracavity photon number using the RESE model (eq. (12) in the main paper) computed for a microlaser ($\beta = 10^{-4}$). The green line shows the line superimposed onto the photon number: its intersection is expected to estimate the threshold value. The value comes close ($C = 0.98$ instead of $C = 1$) but, contrary to expectations, fails to match the GLT.

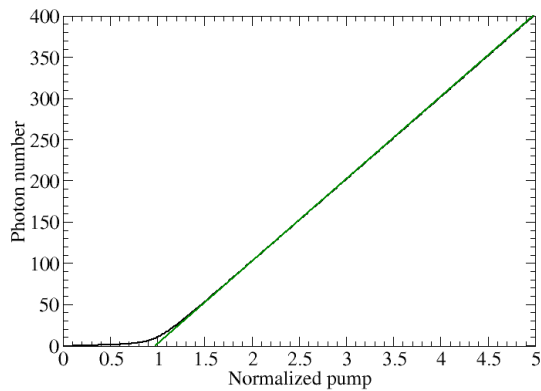

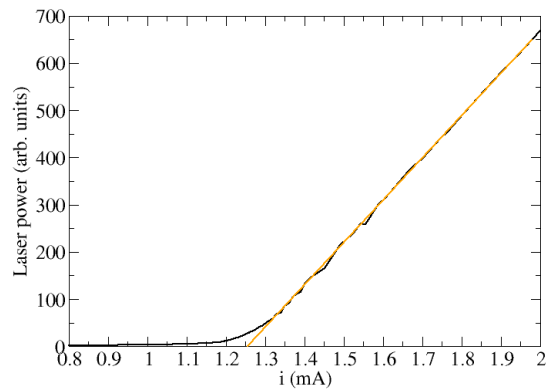

FIG. S-1. Empirical determination of the GLT obtained by extrapolating the photon number growth with normalized pump at large values (straight line) towards the intersection with the 0 photon axis (straight line, green online). The numerical value in this case is 0.98, close to the expected 1. Noise and other perturbations in experiments render the determination much less accurate.

FIG. S-2. Empirical determination of the GLT obtained by extrapolating the photon number growth as a function of current from the measured output of the microlaser (cf. Fig. 2, in main paper). The orange line is superposed onto the laser emission and intersects the 0 value of the laser output power at $i \approx 1.25 mA$. Cf. text for more details.

Repeating the same procedure with the experimental data (cf. Section IV, and in particular Fig. 2, in the main paper) we obtain the result shown in Fig. S-2. The orange line, superimposed onto the laser emission, predicts a threshold value at $i \approx 1.25 mA$, which falls short of the estimated GLT (at $i_{GLT} \approx 1.36 mA$, thus about 8% below the experimental determination of Section IV.B – main paper). Notice that by cutting the pump axis at $i = 2.0 mA$ we have avoided the problem of dealing with the "dent" in the experimental laser response (compare Fig. S-2 to Fig. 2, main paper). Its presence, in one form or another, is well-known in semiconductor lasers and is attributed to saturation effects. While unimportant in the present discussion, aimed at using the figures as illustration, we point out the difficulty in obtaining a reasonable estimate of threshold using this technique. The value can only be approximated, since depending on the range of current used, the best straight line will change, thus affecting (possibly in a considerable way) the determination of threshold.

It is also important to remark that the experimental data are quite noise free, as they have been taken in a microlaser, with a low-noise detector with a long integration time (seconds). This represents an ideal case, which is nearly impossible to match with nanolasers, where the technical difficulties are much greater. Thus, one has to expect a larger degree of uncertainty in the determination of threshold, without, in principle, being able to estimate the actual deviation from a "true" value.

Finally, it is interesting to notice that the laser manufacturer indicates a typical threshold at $i \approx 2.2 mA$, with a maximum value at $3 mA$. These values are based on the need to guarantee a sufficient degree of coherence in the laser output, something that has been shown in the main paper to take place at current values well beyond $i_{GLT}$. Thus, the use of this technique cannot be expected to provide any significant results for microlasers, and, *a fortiori* for nanolasers.

It is important to remark that the critique about the saturation in the laser output should apply to the technique used in Section IV.B (main paper). There, however, help comes from the use of the double-logarithmic scale. Since the lower branch, corresponding to the spontaneous emission, is not affected by saturation effects, and given that the two branches are expected to be parallel, it is easier to determine a reasonable pair of straight lines for the determination of the GLT (as in Section IV.B, main paper).

### B. On the experimental usefulness of threshold indicators proposed in [1]

For the computation of the indicators proposed in [1], based on the derivatives of the laser response function and discussed in Section IV.G of the main paper, we introduce the following functional dependences:

$$f_1 = \frac{d\Phi}{di} \,, \tag{S-1}$$

$$f_2 = \frac{d^2\Phi}{di^2} \,, \tag{S-2}$$

where $\Phi$ represents the output photon flux and takes the place of $L$ in [1], while $i$ is the pump current which replaces $I$ [1]. As shown in Fig. 2 [1], the maximum of $f_2$ can be used as a definition of laser threshold. An alternative, more sensitive definition, is based on the logarithmic derivatives

$$l_1 = \frac{d\log(\frac{\Phi}{1mV})}{d\log(\frac{i}{1mA})} \,, \tag{S-3}$$

$$l_2 = \frac{d^2\log(\frac{\Phi}{1mV})}{d\log^2(\frac{i}{1mA})} \,, \tag{S-4}$$

where the logarithms are in base 10 and the writing $d\log^2(\frac{i}{1mA})$ stands for the second-order differential of the pump's logarithm for the computation of the second derivative. Since – unlike in theory[1] [1] – we are working with dimensional experimental data, the physical quantities have been normalized to convenient values, explicitly indicated in the definitions.

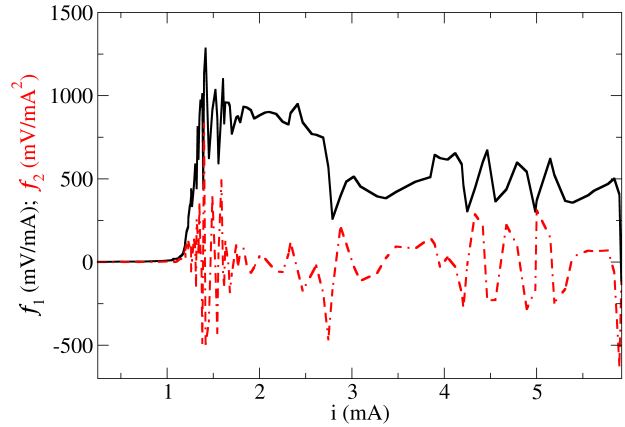

FIG. S-3. Experimental first-order, $f_1$ (solid black line), and second order, $f_2$ (double-dash-dotted red line), derivatives of the laser power as a function of the pump current, computed according to the numerical scheme explained in the text.

We now apply the definitions to the experimental data collected from our micro-VCSEL to test their practical applicability. The first and second-order derivatives are computed from the experimental data according to the following numerical scheme based on the two-column matrix of measured points $(i_j, \Phi_j)$ ($j$ is the index running on all the measurements with $M = max(j) = 122$):

$$f_{1,j} = \frac{\Phi_{j+1} - \Phi_{j-1}}{i_{j+1} - i_{j-1}} \,, \tag{S-5}$$

$$f_{2,k} = \frac{f_{1,k+1} - f_{1,k-1}}{i_{k+1} - i_{k-1}} \,, \tag{S-6}$$

i.e., assigning the numerical differential to the mid-point. $j$ and $k$ run over the whole vector. It is obvious that $f_1$ possesses $M - 2$ point and $f_2$ only $M - 4$. The graphs are plotted as a function of $i(3 \ldots M - 2)$ and run on all points

―――――――

[1] The variables $L$ and $I$ are implicitly assumed to be normalized in [1], otherwise the logarithm could not be taken.

for $f_2$ and only on $f_1(2, M - 1)$, for consistency. Similarly, the logarithmic derivatives are obtained from the numerical definitions, eqs. (S-5,S-6), by simple replacement of $f_{1,2}$ with $l_{1,2}$.

In the main paper, we have shown the computation of the logarithmic derivatives (Fig. 8). Here, Fig. S-3 shows the direct derivatives $f_m$ $(m = 1, 2)$ of the laser response (Fig. 2 in the main paper). Comparing to the nice functional dependences shown in the middle panel of Fig. 2 of Ref. [1], we remark for the first derivative $f_1$:

a. the presence of a large amount of noise, especially in the functional growth of $f_1$ at $i$ values just above $1 mA$;

b. the presence of additional noise in the plateau which should be reached for $i \gtrsim 2mA$, coupled to a reduction in the functional value (expected to be constant).

Not unexpectedly, the situation worsens when considering the second derivative, $f_2$, since noise is further amplified by the second differentiation (double-dash–dotted red line, Fig. S-3). The smooth, clearly peaked curve expected from models (Fig. 2, bottom panel in [1]) appears as a noisy oscillation around 0, with no hint of any usable information. It is important to remark that threshold should have been identified by $\frac{df_2}{di} = 0$ (horizontal tangent at maximum), which is replaced here by an ensemble of numerous maxima, whose most probable origin is the intrinsic noise. Yet, inspection of the original curve (Fig. 2 in the main paper) would not have anticipated such a poor outcome, since the experimental data appear to be sufficiently smooth.

### 1. Noise filtering of the experimental data

The presence of noise in experimental data comes as no surprise, even when their quality is already high. It is therefore important to check whether adequate noise removal can sufficiently palliate the difficulties encountered in the previous subsection before abandoning the proposed scheme [1].

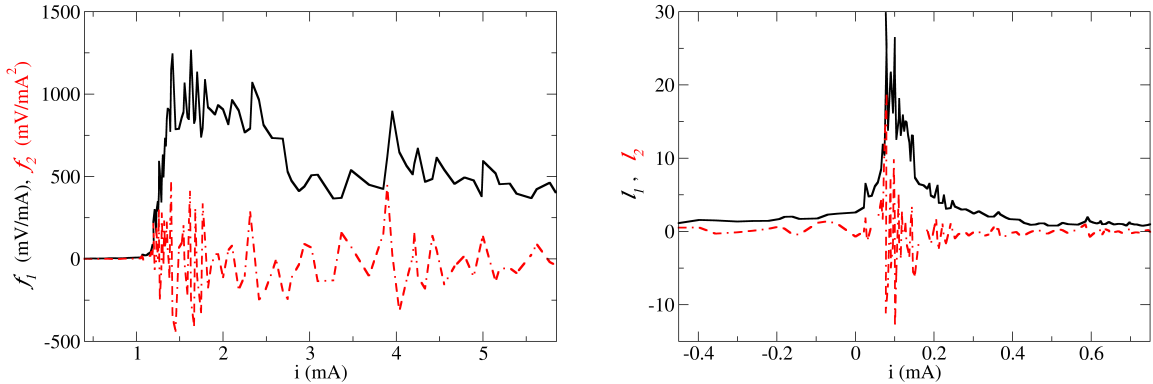

FIG. S-4. Left panel: first $f_1$ and second $f_2$ derivatives of the laser response function (Fig. 2, main paper) computed on the smoothed dataset $\tilde{\Phi}$. Right panel: logarithmic first $l_1$ and second $l_2$ derivatives, again for the same dataset.

Smoothing over a 5-point algorithm with exponential weights, in the form

$$\tilde{\Phi}_j = \frac{\sum_{i=j-2}^{j+2} \Phi_i e^{-|i-j|}}{\sum_{i=j-2}^{j+2} e^{-|i-j|}}, \tag{S-7}$$

reduces the amount of noise present on the laser input-output curve measured in the experiment. Applying the derivative algorithms to the $\tilde{\Phi}$ dataset provides the smoothed estimates of the derivatives shown in Fig. S-4, where the italicized quantities ($f$ and $l$) represent the derivatives of the smoothed dataset $\tilde{\Phi}$. Comparison between the derivatives computed from the raw experimental data and from the smoothed ones shows that there is very little improvement, if any.

While this does not exclude that more sophisticated smoothing techniques may improve a bit the functional form of the derivatives, it is unlikely that without excessive data manipulation the derivative predictors may provide significant information. The more processing is applied to the data, the more likely it is that the information contained in the data may be distorted, providing indicators whose predictions depend more on the processing procedure than on the information retrieved from the experiment. Furthermore, as already stressed previously, these poor results are obtained from a microlaser experiment, where noise is still manageable and large amount of data can be collected with relative ease. The stronger influence of noise on the very low photon flux obtained from nanolasers is almost certainly going to worsen the quality of the predictions obtained from the derivative-based indicators.

We must therefore conclude that, although the proposed definitions [1] are sound and grounded in fundamental physical features, they do not hold any practical usefulness for the identification of the laser threshold.

### C. RIN

In order to complete the picture of the information coming from the experiment, we compute the Relative Intensity Noise from the data, as done for the Fano factor. Notice that the results can be obtained either from the variance and the square of the photon average, or from the Fano factor and the photon average (eq. (2) in main paper).

The result is shown in Fig. S-5, which, again, does not show any particularly useful information. The expected plateau at low pump (Fig. S-22) is not visible. Its presence stems from the noisy emission below threshold *devoid of photon bursts*. Their presence enhances the fluctuations and therefore the noise, making the shoulder disappear at low pump. Thus, even the RIN cannot provide any useful information on the threshold location.

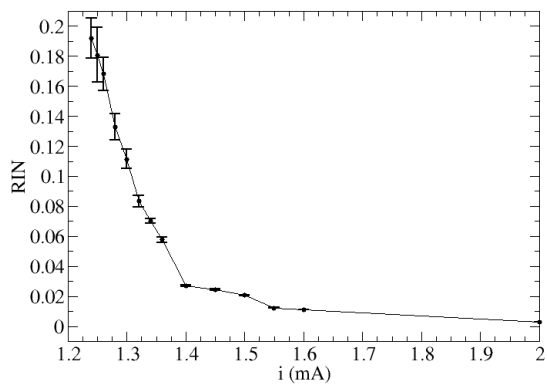

FIG. S-5. Relative Intensity Noise (RIN) computed from the experimental data.

### D. How to experimentally identify the Bifurcation Threshold

In the main paper, we have shown the fallacy of the secondary threshold definitions for non-macroscopic lasers and stated that the Bifurcation Threshold (BT) is the only one whose validity is not influenced by the system size. While several techniques are in use to identify the other thresholds – and we have used some of them – the BT is not often sought, thus it is useful to focus on a couple of experimental possibilities. The selection is somewhat arbitrary and not exhaustive; other techniques may be developed, but this small overview helps offering an idea of how to proceed (especially for the benefit of non-experimentalists).

#### 1. Interferometric approach

The BT is based on the emergence of a *coherent electromagnetic field* out of noise. Thus, we need to measure the field, rather than its intensity which is non-zero also for the spontaneous emission (whose field instead has zero average). One of the immediate possibilities to measure an optical field is to make it interfere, so as to measure the resulting intensity with a response that is nonlinear in phase. Indeed, in the optical region – unlike in the radiofrequency domain – there are no detectors capable of directly measuring a field strength.

Single-beam interferometers (Michelson or Mach-Zehnder) are the most direct answer to the quest (Fig. S-6). They rely on the self-interference of a single beam which is split into two parts. The interferometers can be adjusted to the same exact length by obtaining a signal onto the detector with white light (or spontaneous emission), thus the name "white light fringe". Then the length of the variable arm is gradually moved away from the equilibrium position. In the absence of a coherent field, the signal goes rapidly to zero (typically within a couple of hundred nanometer displacement), while if a coherent field exists, then the signal persists and decays away more slowly, giving a measure of the coherence length (and time) for the signal. This is exactly the technique used to measure $g^{(1)}(\tau)$ (and from it reconstruct the spectral features of the optical signal through Fourier transform) [2].

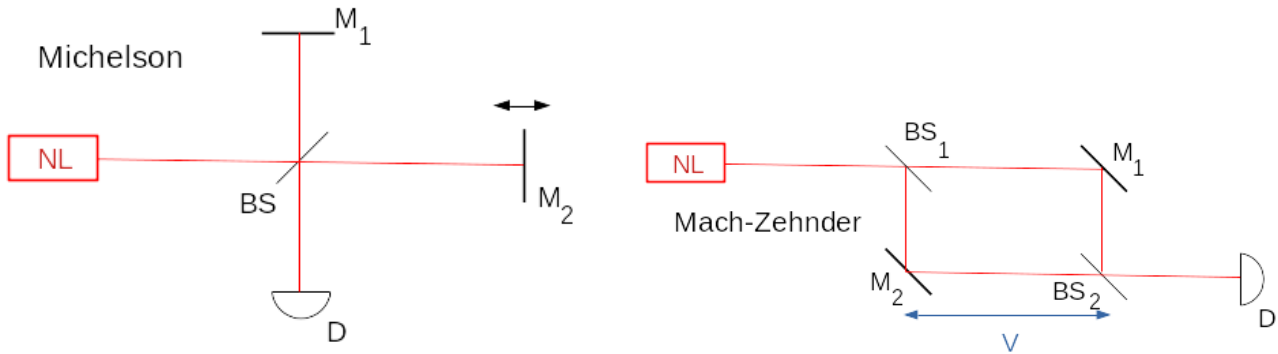

FIG. S-6. Schematics of principle of two interferometric set-ups for the (coherent) field amplitude measurement. For the Michelson interferometer (left) the nanolaser emits light beam split into two equal portions by the beam splitter ($BS$), and reflected back onto detector $D$ by the two mirrors ($M_1$ and $M_2$). Mirror $M_2$ can be moved with sub-micrometric precision to control the path difference, thus the interference between the two beams. For the Mach-Zehnder interferometer (right) the beam is first split by $BS_1$ into two equal beams which are reflected by mirrors $M_1$ and $M_2$, respectively, to be recombined onto detector $D$ by $BS_2$. The double-arrow along the bottom path (denoted $v$) identifies the variable length of the bottom arm (to be properly implemented) which enables the same operation done with the Michelson interferometer.

Thus, the identification of the BT relies on the detection of a signal which emerges from detector $D$ with sufficiently large correlation length. Given the weakness of the signals, the measurement is not trivial. A practical implementation can be best obtained in a fibred system, but other technical problems need to be solved, such as maintaining the polarization along the whole path (so as to not destroy the interference signal) and the availability of suitable mirrors (or "circulators") at the correct wavelength [3]. The possibility of using Michelson or Mach-Zehnder interferometers adds a bit of freedom to reduce the technical constraints imposed by the component availability.

### 2. Signal heterodyne

Rather than splitting the signal into two parts, which amounts to a homodyne detection, we can turn to heterodyning the signal emitted by the laser with that of a so-called *local oscillator*: another laser which is frequency- and amplitude-stabilized to emit a reference signal to which the output of the nanolaser is compared. The schematics of principle is much simpler here, since it suffices to combine the two laser outputs onto detector $D$ through a beamsplitter, $BS$ (Fig. S-7).

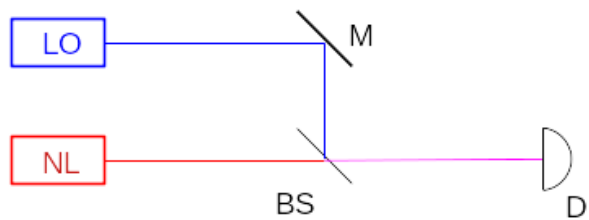

FIG. S-7. Schematics of principle of the heterodyne measurement for the identification of the BT. LO is the Local Oscillator (tunable stabilized laser), NL the nanolaser whose BT is measured. The rest of the components are the same as in Fig. S-6.

The complexity is transferred onto the performance of the Local Oscillator, which has to be tunable – to match the nanolaser emission wavelength –, in addition to possess a very good frequency and amplitude stability. The measurement is done by finding the emission frequency of the nanolaser (above threshold), then detuning the LO by a few GHz to obtain a beat note (frequency difference between the nanolaser and the LO), finally changing the nanolaser pump to find the value for which the beat note emerges. The beat note measurement is best done with a fast (but sensitive) detector (bandwidth exceeding $10\,GHz$) and with a radiofrequency analyser. The latter needs to have a good frequency range which matches the detector's in order to allow for a good "capture" frequency range (but typically rf analysers have a range up to $50GHz$ and are not the limiting element).

## S-2. COMPLEMENTS ON THE RATE EQUATIONS (THERMODYNAMIC LIMIT)

The beauty of the Rate Equations is that they can be phenomenologically derived from the fundamental physical processes which connect the radiative transitions between atomic levels (or bands in solids and semiconductors) on the basis of absorption, stimulated and spontaneous emission [4]. This provides a direct and powerful physical insight into the interaction between radiation and matter and the transition between an incoherent and a coherent field emission. The same identical equations can be obtained from a microscopic approach where the physics of a two-level atom is described with Bloch equations [5] and the (typically classical) electromagnetic field through Maxwell equations [6]. The adiabatic elimination of the atomic polarization for a single cavity mode laser leads to the REs (the electromagnetic field phase decouples from the description).

The most relevant pieces of information have been given in the main paper. It is useful, however, to dispose of a few other for comparison with the RESE. For the RESE eigenvectors, close to threshold, please refer to [7, 8].

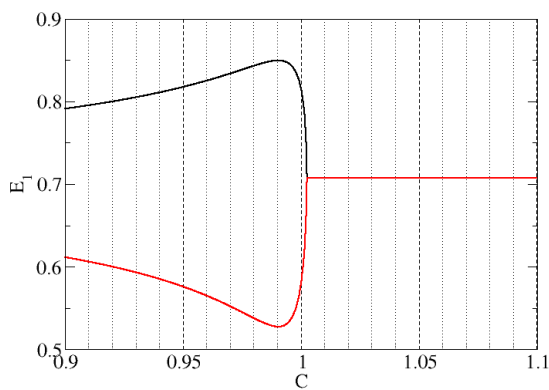

FIG. S-8. Components of the first normalized eigenvector, $E_1$, corresponding to ${}^l\lambda_{t,1}$. The black line matches the $N_t$ direction, while the red one corresponds to $n_t$ direction. Below threshold this eigenvector is predominantly aligned with the $N_t$ component. Only above threshold do the two components approach, to become equal (alignment along the plane diagonal) once the two eigenvalues ${}^l\lambda_{t,1}$ and ${}^l\lambda_{t,2}$ become complex conjugate. Notice that at $C = 1$ the black component still dominates.

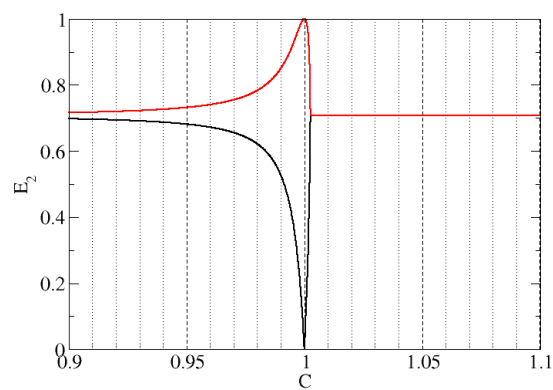

FIG. S-9. Components of the second normalized eigenvector, $E_2$, corresponding to ${}^l\lambda_{t,2}$. The black line matches the $N_t$ direction, while the red one corresponds to $n_t$ direction. Below threshold this eigenvector is aligned along the bisectrix of the plane, but rapidly aligns with the $n_t$ component. At $C = 1$ it is entirely aligned along the photon component to then return to the bisectrix once the two eigenvalues ${}^l\lambda_{t,1}$ and ${}^l\lambda_{t,2}$ become complex conjugate.

Some additional information can be gained from the eigenvectors. Fig. S-8 shows the moduli of the two components of the eigenvector $E_1$, aligned with the first eigenvalue ${}^l\lambda_{t,1}$, which remains stable (${}^l\lambda_{t,1} < 0$) in the whole interval (solid line in Fig. 15, main paper). The graph shows a detail of the interesting region (close to threshold). The carrier component, aligned along $N_t$, dominates in this eigenvalue especially just before threshold. Fluctuations along this eigendirection (mostly aligned with the carriers) decay towards the stable solution.

Fig. S-9 gives the same information for the other eigenvalue, ${}^l\lambda_{t,2}$, which changes stability at $C = 1$ (dashed line in Fig. 15, main paper). While initially aligned along the plane bisectrix in the $(N_t, n_t)$ representation, this eigenvalue acquires a dominant photon component (red line) which becomes the only non-zero one at $C = 1$. As for $E_1$, $E_2$ aligns again along the bisectrix once the eigenvalues become complex conjugate.

The interesting information gathered from the eigenvectors is that the fluctuations grow in the direction of the photon axis, as one could intuitively expect. However, this is a peculiarity of the REs in the thermodynamic limit, as mainly shown in [8]. Since we are looking only at the moduli of the two eigenvector components, we cannot distinguish the angle which $E_1$ and $E_2$ form with the reference system. This information can be gathered looking at the real parts of the normalized eigenvalues, where we see that there are rapid rotations (Fig. S-10). The imaginary components are also given, for completeness.

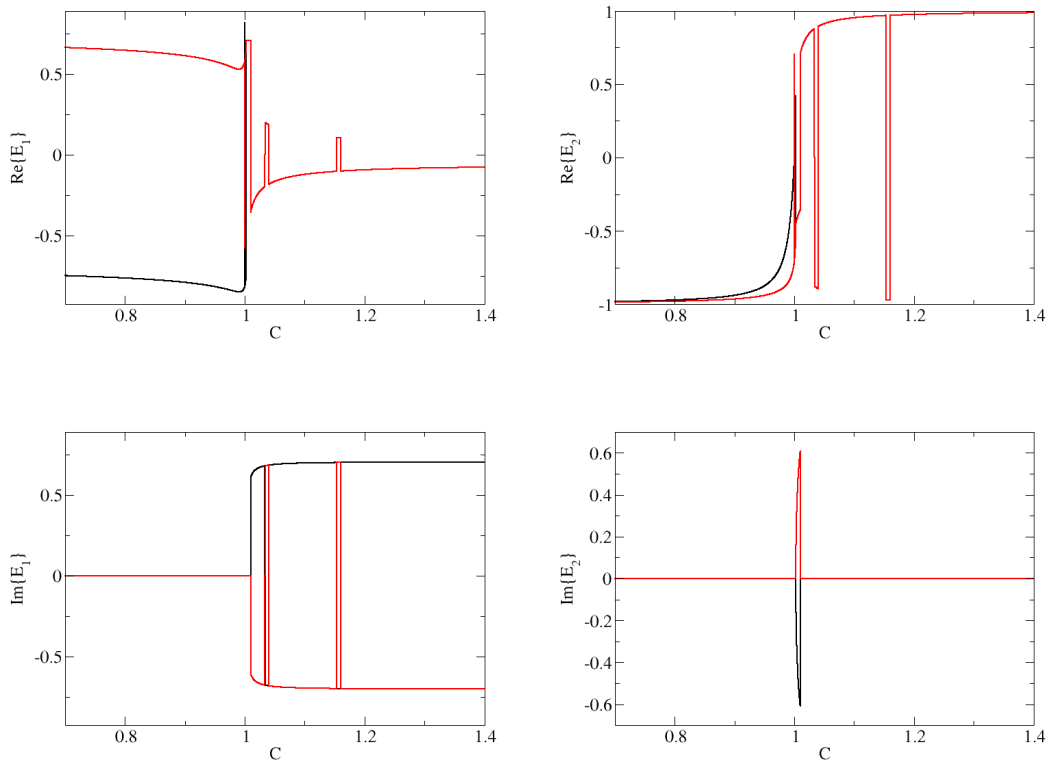

FIG. S-10. Real and imaginary parts of $E_1$ and $E_2$. The color coding follows that of Figs. S-8 and S-9. Where only the red is shown, the black line is exactly superposed.

## S-3. COMPLEMENTS ON THE RATE EQUATIONS WITH SPONTANEOUS EMISSION

The main paper presents the main features of the Rate Equations with Spontaneous Emission (RESE). Here, we examine some pieces of complementary information which are useful for a more complete overview, or for comparison with the experimental results (when $\beta = 10^{-4}$, a value fairly close to that of the micro-VCSEL).

### A. Laser input-output characteristics

The $\langle n \rangle$ steady state curves (laser response) for the physical solution (positive photon number, only) for all $\beta$ values are displayed in Fig. S-11, where the pump rate is expressed in cycles per second. This classic picture (e.g., [9, 10]) shows that the asymptotic laser output (above threshold) is the same irrespective on the value of $\beta$, as long as the pump rate can be made large enough (of course, this is unfeasible in experiments, since there is a practical limitation to the current density in a device, which undergoes damage beyond a construction-specific value). The second interesting observation is that average number of photons – described by the fraction $\beta$ of the spontaneous emission coupled into the lasing mode – is practically the same for all lasers, as well (as long as $\beta$ is small enough, approximately $< 10^{-2}$, as clear in Fig. S-11). Thus, the overall behavior of the different lasers is the same in the two limits: far below and far above threshold. What changes as a function of $\beta$ (i.e., approximately as a function of electromagnetic cavity volume) is the shape of the transition between the two branches. It is also important to remark (although difficult to see in the figure) that the slopes of the two oblique branches is the same.

Finally, as already mentioned in the main paper, we have neglected non-radiative relaxation processes (accounted for in [10], for instance) which lend a "curvature" to the $\beta = 1$ laser characteristics (black line in Fig. S-11). This choice is made to illustrate the physical principles of the influence of laser scaling, rather than to model specific devices.

The same response, normalized to each corresponding threshold value, as defined in [11], better displays the increasing sharpness of the jump and illustrates the role of the $\frac{1}{\sqrt{\beta}}$ mid-point as the GLT (cf. main paper for the definition of GLT). The threshold normalization is given by eq. (14) in the main paper and straightforwardly provides eq. (12).

Displayed with the pump rate in normalized form, Fig. S-12, the laser response highlights the sharpness of the jump between the lower (incoherent) and upper (coherent) branch of emission and allows for a better visual positioning of the GLT (mid-height point in the "jump" for each curve).

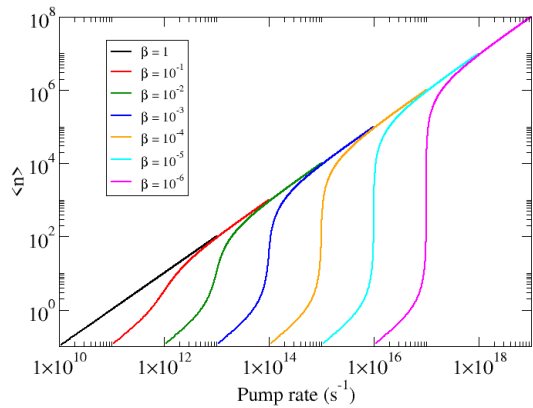
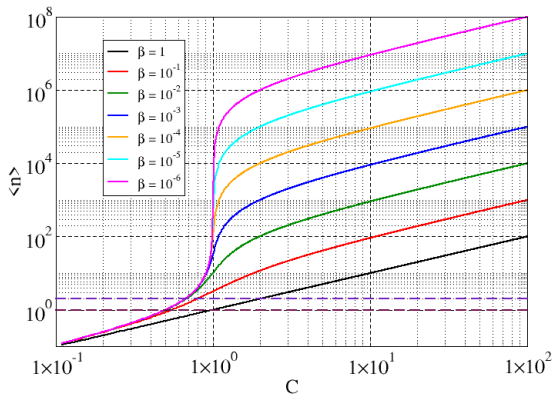

FIG. S-11. Average photon number $\langle n \rangle$ for the physical solution plotted as a function of pump rate for the $\beta$ values indicated in the figure. Notice that the upper branch converges to the same value for all kinds of lasers, while the threshold is shifted away towards larger pump as $\beta$ decreases. These curves are plotted using eq. (12) from the main paper, undoing the threshold normalization with the help of eqs. (14).

FIG. S-12. Average photon number $\langle n \rangle$ for the physical solution plotted as a function of normalized pump for the $\beta$ values indicated in the figure. The mid-point in each curve corresponds to the $\frac{1}{\sqrt{\beta}}$ [11], thus the GLT.

### 1. Comments on the Quantum Threshold

The horizontal dashed line (indigo, Fig. S-12) shows the $\langle n \rangle = 2$ value, proposed as a better approximation of the Quantum Threshold, QT [10, 12]. Using considerations of principle, one should choose $\langle n \rangle = 1$ (maroon long-dashed line) for the QT, but it is sometimes deemed preferable to consider the average photon value where there is, in average, one stimulated plus one spontaneous photon [10, 12]. The definition of the QT is not univocal and, in spite of its conceptual appeal, has been criticized in [11] for its intrinsic ambiguity.

Regardless of the criticism, we can still extract three remarks on the QT:

1. The predicted QT, according to the RESE, does not coincide with the GLT and spans from $0.68 \leq C_{QT} \leq 2$.

2. QT would coincide with the the GLT, but only for the $\beta = 1$ laser, if one were to take the cut at $\langle n \rangle = 1$. Clearly, this definition is not satisfactory, either, since it is correct only for the "thresholdless" case; as such it lacks generality. In order to recover generality, we need to use the construction applied in the main paper (Fig. 1).

3. The intersection between the horizontal line and the laser response rapidly converges onto one point when considering sufficiently large lasers ($\beta \lesssim 10^{-3}$), thus showing the tendency towards a generic behavior as the laser gains in size. However, the apparent convergence from $\beta \approx 10^{-3}$ is only a visual indicator, belied by a more thorough analysis (Sections VII and VIII in the main paper and Section S-4).

### B. Eigenvalues of the nonphysical solution

For the REs, i.e., in the thermodynamic limit, we have shown the stability of both solutions. In the RESE, only one has physical meaning, since the other one corresponds to $\langle n_- \rangle < 0$. For completeness, we show its eigenvalues (Fig. S-13). The presence of one positive eigenvalue ($\lambda_1$ in the figure) for all pump values shows that this nonphysical solution is always unstable. This ensures that only the physical solution is accessible for all pump values. The figure has been computed for $\beta = 10^{-4}$, but the qualitative result remains the same for all $\beta$ values.

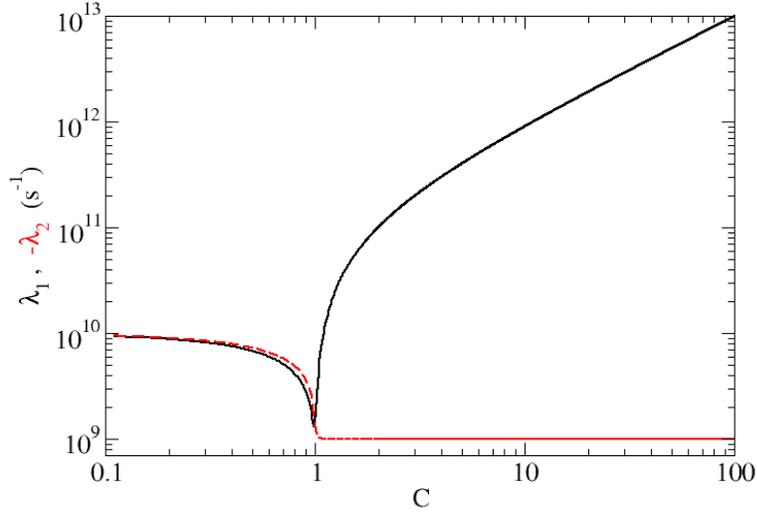

FIG. S-13. Eigenvalues for the nonphysical, $\langle n_- \rangle$, solution computed for $\beta = 10^{-4}$, as a function of $C$. Both are strictly real over the whole interval. $\lambda_1 > 0$ (solid black line) for all values of $C$, while $\lambda_2 < 0$ (dashed red line). Notice that the graph plots $-\lambda_2$ in order to use a logarithmic representation.

## S-4. THRESHOLD PREDICTIONS WITH RESE

The main features of the RESE have been analysed in the main paper (Sections VII and VIII). The following sections present specific aspects, mainly related to the predictions that the model offers on the basis of the various threshold definitions. This material represents a parallel to the analysis of the various threshold definitions considered in the experimental section of the main paper (Section IV). Here, we separately consider, when suited, a fraction of spontaneous emission coupled into the lasing mode ($\beta$) close to the experimental one ($\beta = 10^{-4}$), and extend the analysis to the full range of $\beta$ values.

### A. Gain-Loss Threshold from RESE

As already specified in the main paper (Sections III.A and IV.B), the GLT is identified by the pump value for which $C = 1$ (eq. (14) in main paper), where the photon number becomes $\langle n \rangle_{C=1} = \beta^{-1/2}$ [11]. As such, it corresponds to the point at mid-height in the laser response curve (dot in Fig. 16 of main paper) for all values of $\beta$. Given the clear mathematical origin of the definition within the context of the RESE, this is an easy point to identify for all values of $\beta$, as mentioned in Section S-3 A with reference to Fig. S-11.

### B. Photon Statistical Threshold (PST) from RESE

The photon statistical threshold can be computed from the model estimating the variance of the photon number (in response to a fluctuation) and using the expression for the second order, zero delay autocorrelation (eq. (3) in the main paper) which can be reformulated in a more convenient form[2]:

$$g^{(2)}(0) = 1 + \frac{\langle \Delta n^2 \rangle}{\langle n \rangle^2} , \tag{S-8}$$

where $\langle \Delta n^2 \rangle$ stands for the variance of the photon number. The latter[3] can be computed from a Langevin representation of the stochastic forces in the linearised form of the RESE [14, 15], neglecting the noise force in the carrier rate

---

[2] We are basing this expression on the definition given in eq. (3) in the main paper, for $\tau = 0$, rather than on the quantum definition $g^{(2)}(0) = \frac{\langle n(n-1) \rangle}{\langle n^2 \rangle}$, useful for very weak photon fluxes. This is due to the fact that the number of photons that our detector collects is always macroscopic [13]

[3] We are grateful to J. Mørk for the computation of the explicit expression.

equation – a very satisfactory approximation over a broad pump range [16]:

$$\langle \Delta n^2 \rangle \approx \frac{\gamma(1+\beta n)n(n+1)}{\frac{\Gamma_c}{n+1} + \gamma(1+\beta n)} \left(1 + \frac{[\gamma(1+\beta n)]^2}{(\beta \gamma \Gamma_c n) + \gamma(1+\beta n)\frac{\Gamma_c}{n+1}}\right), \tag{S-9}$$

where all parameters have been defined in Section VI of the main paper.

### C. Microlaser PST from RESE

Fig. S-14 shows the computed second-order, zero-delay autocorrelation function for $\beta = 10^{-4}$, thus for a $\beta$ value close to the experimental one. The autocorrelation decreases monotonically from 2, with a deviation which starts at $C \approx 0.97$, thus before the GLT. The anticipated reduction is part of the contribution of the spontaneous emission, in RESE, to the coherent field which raises the photon number (and reduces fluctuations) before $C = 1$. However, none of the structure observed in the experiment is visible here, as the dynamics reported in Section IV.H (main paper) is only partially captured by the RESE.

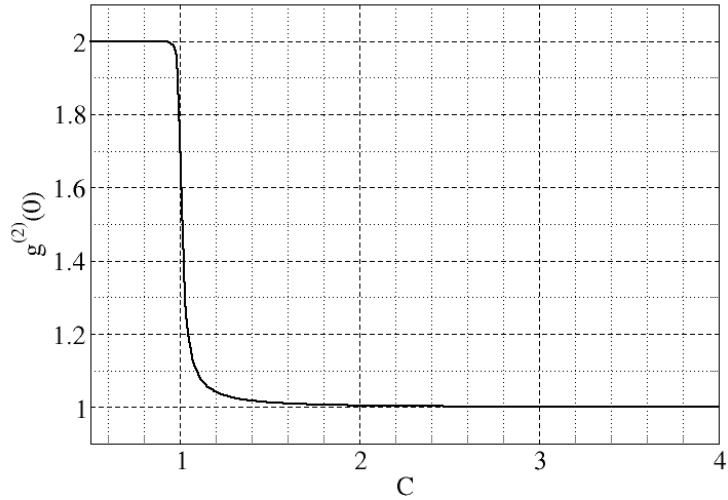

FIG. S-14. Second-order zero-delay autocorrelation function for $\beta = 10^{-4}$, computed from RESE.

Convergence towards Poisson statistics does not occur at $C = 1$, unlike the thermodynamic limit, but a deviation of only 1% ($g^{(2)}(0) = 1.01$) is observed at $C \approx 1.6$, with 0.2% at $C \approx 3$. Again, these values do not agree with the experimental observations, since in the experiment $g^{(2)}(0) \approx 1.3$ approximately 60% above the GLT (cf. Fig. 6 in [13] for the experimental autocorrelation function devoid of the bias introduced by the limited bandwidth of the data acquisition chain). Aside from quantitative disagreements, however, Fig. S-14 confirms that the PST does not coincide with the GLT in microlasers.

### D. General PST considerations

Application of the same procedure to the estimate of $g^{(2)}(0)$ for different $\beta$ values produces the curves displayed in Fig. S-15. The functional form expected for macroscopic lasers is observed (black line): a sharp drop from $g^{(2)}(0) = 2$ to $g^{(2)}(0) = 1$ at $C = 1$, signaling the transit from incoherent to coherent emission. Approaching the limit between macro- and mesoscopic lasers (e.g., $\beta = 10^{-5}$, green line) a rounding off of the transition is visible, providing a delayed convergence towards Poisson statistics already for the larger mesolasers ($\beta = 10^{-4}$, blue curve). The details of the progressively smoother convergence towards $g^{(2)}(0) = 1$ are best seen from Fig. S-16. As the laser size is further reduced (i.e., $\beta$ growing larger) the transition becomes progressively smoother, with the Poisson limit reached only for $C > 100$ for the largest $\beta$ value ($\beta = 1$, purple – cf. Fig. S-15).

A second aspect, already remarked upon in Section S-4 C, concerns the shape of the autocorrelation for $C < 1$. The deviation from the sharp drop, observed in Fig. S-14 become more pronounced as $\beta$ grows, with a non completely thermal statistics below threshold (cf., e.g., $g^{(2)}(0)$ for $C < 1$ in the $\beta = 1$ case). Subthermal statistics is predicted below threshold also in Quantum-Dot-based devices [17] for large $\beta$ values, adding a physical aspect that combines with the well-known difficulty in obtaining a good experimental measurement of thermal statistics in below-threshold nanolasers [18]. We remark, however, that the RESE, like all analytical models developed so far, does not account for the experimentally observed [13, 19] and stochastically predicted [19] superthermal statistics.

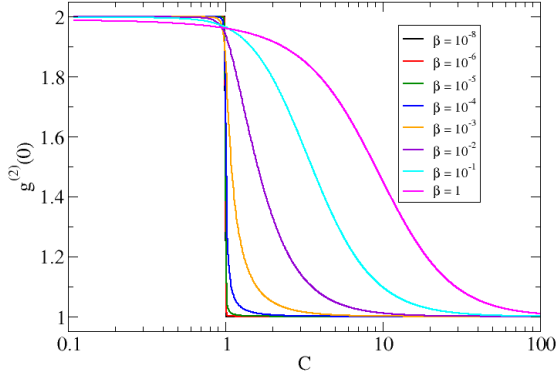

FIG. S-15. Second-order zero-delay autocorrelation function for lasers with different fraction of spontaneous emission coupled into the lasing mode computed from RESE. The legend identifies the different curves.

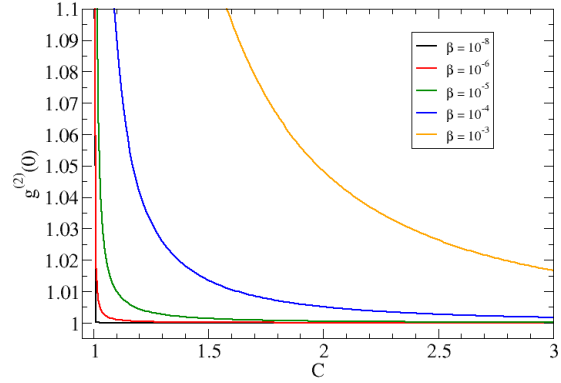

FIG. S-16. Detail of the convergence towards Poisson statistics for the second-order zero-delay autocorrelation function for micro and macrolasers (computed from RESE).

The predicted shape of $g^{(2)}(0)$, for the different $\beta$ values, matches the reported shapes for the Fano factor (Section VIII.A, main paper). However, it also highlights a contraction that emerges from these two different *threshold indicators*. Focusing on $\beta = 1$, we observe that the maximum in the Fano factor, $F_{max}$, corresponds to $g^{(2)}(0) \approx 1.5$ – hardly an indication of field coherence[4]!

We already mentioned, in the main paper, that the autocorrelation is not to be considered as a reliable threshold indicator for nanolasers, but self-consistency should be required when passing from one threshold definition to the other (FT and PST), if we want to consider them as meaningful! This discrepancy is an additional confirmation of the fact that the accessory threshold definitions are meaningful only in the thermodynamic limit and can be extended, at most to macroscopic lasers.

### E. Microlaser Fano Threshold from RESE

In order to offer a detailed comparison with the experiment, we trace here the Fano factor (Fig. S-17) for a microlaser with $\beta = 10^{-4}$ (close to that of the micro-VCSEL). The maximum of the Fano function occurs at $C = 1.05$, thus 5% beyond the GLT. A similar discrepancy is observed in the experiment (Section IV.D), but there the FT precedes the GLT, rather than following it. More important is the discrepancy in the shape of the curves, which show a much more complex structure in the experiment than in Fig. S-17. This fact has been amply discussed and originates from the more complex dynamics experimentally observed.

### F. Relaxation Oscillation Threshold from RESE

The ROT can be easily obtained from the *lsa* of RESE (Section VII.B) by looking at the behavior of the imaginary part of the eigenvalues. As soon as the latter become non-zero, the dynamics between photons and carriers shows coupled oscillations. A global picture, in physical pump units, has been offered in Fig. 21 of the main paper.

Fig. S-18 shows in semilogarithmic scale the behavior of $\lambda_i$, displaying a sharp growth out of 0, followed by a rapid saturation. The normalized pump value $C$ at which the ROT occurs changes rapidly and macroscopically as the nanolaser regime is reached. For microlasers, the threshold varies in a smaller range, not easily recognizable in the figure.

---

[4] The example taken here is extreme, but well illustrates the problem, which persists to a lesser extent for all other non-macroscopic lasers.

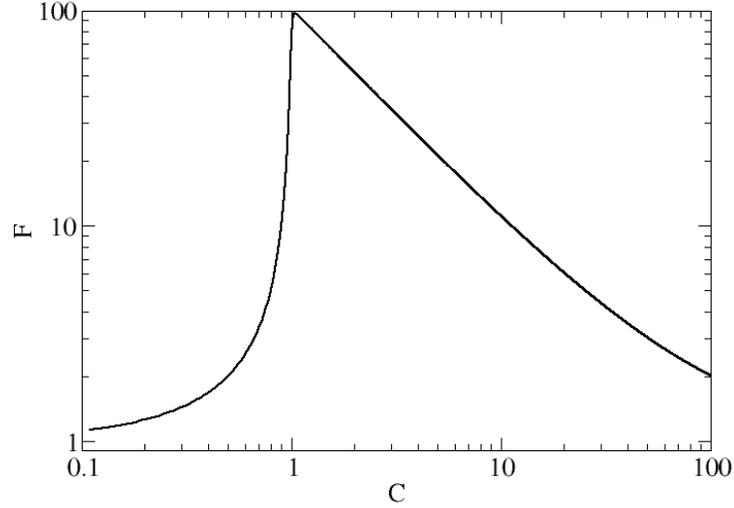

FIG. S-17. Fano function for a $\beta = 10^{-4}$ microlaser, computed from RESE.

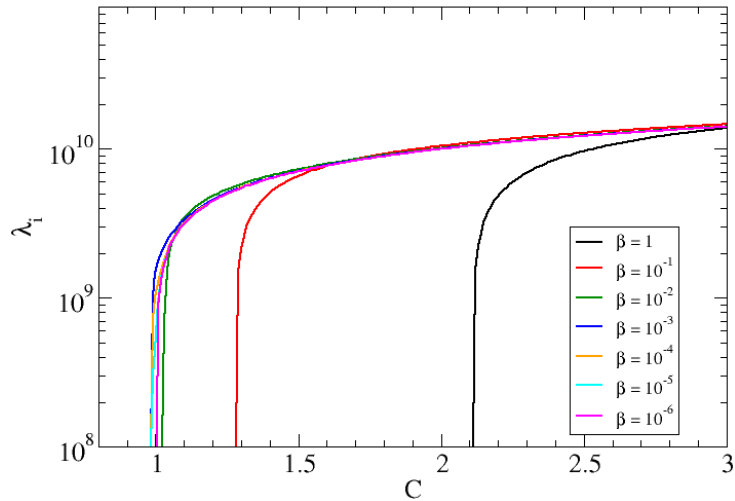

FIG. S-18. Imaginary part of the stability eigenvalues for different $\beta$ (cf. legend) – computed from RESE.

Macroscopic lasers display a much more complex behavior (Fig. S-19), introducing through this change in the evolution of $\lambda_i$, a clear boundary with microlasers. An inflection point appears for $\beta \approx 4 \times 10^{-5}$, from which a minimum in the evolution of $\lambda_i(C)$ develops, leading to a pinching of a first oscillatory region in the shape of a *bubble* for $4 \times 10^{-6} \leq \beta \leq 3 \times 10^{-6}$. The *oscillatory bubble* shrinks as $\beta$ decreases and is entirely absent at $\beta = 10^{-9}$, rejoining the thermodynamic limit (Fig. 23 in main paper).

The complementary picture, which shows the real part of the eigenvalues S-20 under the same conditions, provides additional illustration of the evolution of the dynamics. Here the *bubble* is complementary to the one of the imaginary parts, since the real parts separate and reunite (for $\beta = 3 \times 10^{-6}$ and $4 \times 10-6$) in the region where the imaginary part disappears. The smallest $\beta$ value matches the one chosen to illustrate the convergence towards the thermodynamic limit (Fig. 23 in main paper).

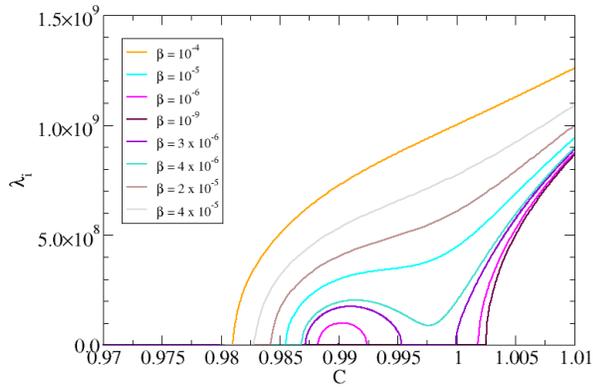
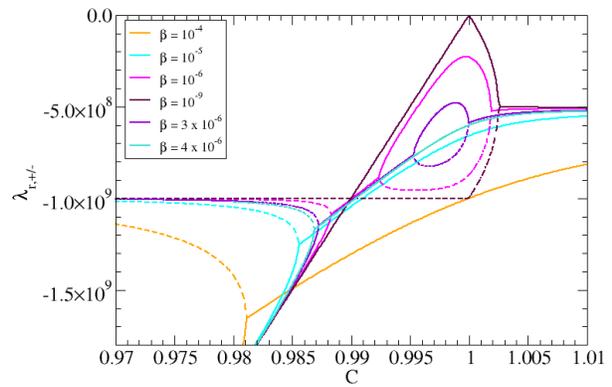

FIG. S-19. Expanded region of the imaginary part of the stability eigenvalues for small $\beta$ values (cf. legend) – computed from RESE.

FIG. S-20. Detail of the real part of the stability eigenvalues for different $\beta$ (cf. legend) – computed from RESE.

### G. Largest Fluctuation Threshold – from RESE

We conclude the comparison with the experimental results by looking at the largest fluctuations (Fig. S-21) and the relative fluctuations (Fig. S-22) predicted by the RESE, compared to their experimental counterparts (Figs. S-23 and S-24, respectively). The comparison displays the hallmarks of the differences coming from the more complex dynamics observed in the micro-VCSEL, confirming that the indicators are not very useful.

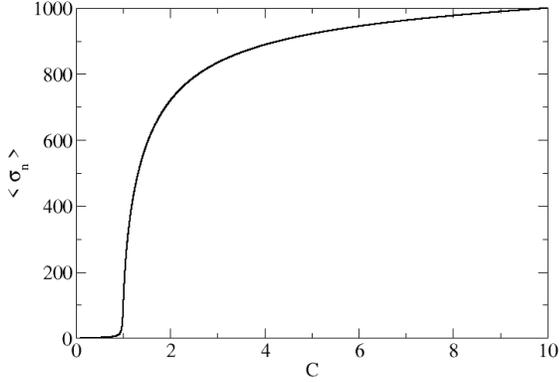

FIG. S-21. Absolute fluctuations computed from RESE.

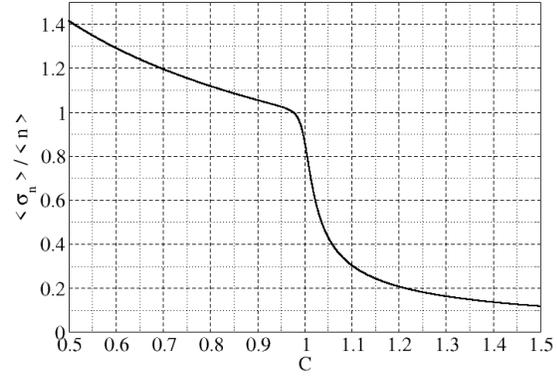

FIG. S-22. Relative fluctuations computed using the variance (eq. (S-9)) and the average photon number for the RESE (eq. (12) in main paper).

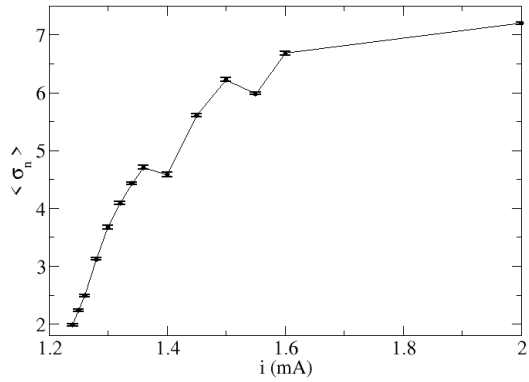

FIG. S-23. Absolute photon fluctuations $\sigma_n$ as a function of pump current. The averages for each displayed point are obtained from 10 data files of $10^5$ points each, thus allowing for the computation of the error bar.

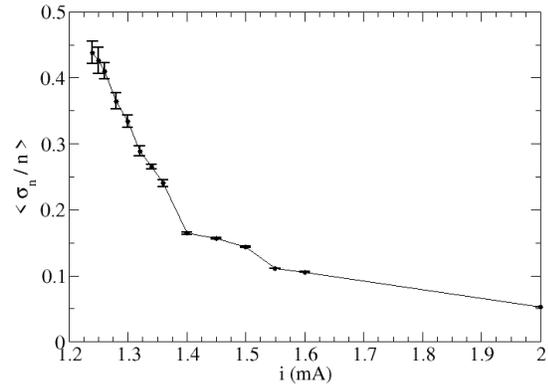

FIG. S-24. Relative photon fluctuations $\frac{\sigma_n}{\langle n \rangle}$ as a function of pump current. The averages for each displayed point are obtained from 10 data files of $10^5$ points each, thus allowing for the computation of the error bar.

## H. On RIN

The Relative Intensity Noise (RIN) is at times preferred for the characterization of lasing, thanks to its direct visualization of the decreasing noise (and its proper normalization, unlike the Fano factor, where the denominator has dimensions equal to the square root of those of the numerator).

Plotting the RIN predictions from the RESE we obtain the results of Fig. S-25, which shows the usual picture for macroscopic lasers: a sharp drop in the noise level followed by a linear decay in double logarithmic scale. The drop is gradually reduced as $\beta$ grows, leading to the "anomalous" growth (matching the maximum in the Fano factor, Fig. 24, main paper), followed by the recovery in decay, for nanolasers. Notice that at large pump values the asymptotic scaling is recovered when the (normalized) pump is sufficiently large.

Comparison between the RIN extracted from the experimental data (Fig. S-5) and the blue curve in Fig. S-25 shows, again, the emergence of structures reminiscent of the more complex experimental dynamics.

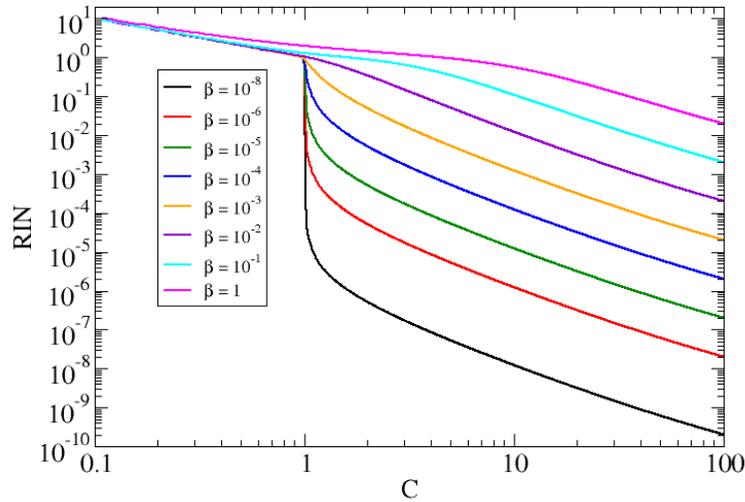

FIG. S-25. Relative Intensity Noise function for lasers with different fraction of spontaneous emission coupled into the lasing mode computed from RESE.